\documentclass[conference]{IEEEtran}
\IEEEoverridecommandlockouts
\usepackage{amsmath,amssymb,amsfonts}
\usepackage{balance}

\usepackage{upgreek} %upgreek used for \wordsize command
\usepackage{mathtools} % for \coloneqq
\usepackage[noend]{algorithmic} %[noend] option omits "end if" and "end for" from algorithms
\usepackage{algorithm}
\usepackage{graphicx}
\usepackage{textcomp}
\usepackage{xcolor}
\def\BibTeX{{\rm B\kern-.05em{\sc i\kern-.025em b}\kern-.08em
    T\kern-.1667em\lower.7ex\hbox{E}\kern-.125emX}}
    
\usepackage[commentmarkup=uwave,draft]{changes}
\usepackage{tablefootnote}

\usepackage{url}

\usepackage{breakurl}
\makeatletter
\let\NAT@parse\undefined
\makeatother
\usepackage[breaklinks,hidelinks,colorlinks=true,linkcolor=blue,citecolor=blue,pagebackref=true]{hyperref}

\usepackage[firstpage=true,color=black]{background}

\SetBgContents{Accepted Manuscript: 2022 IEEE International Symposium on Secure and Private Execution Environment Design}% Set contents
\SetBgPosition{current page.east}% Select location
\SetBgVshift{0.6cm}% Add vertical shift (results in a shift in x direction due to rotation)
\SetBgOpacity{0.5}% Select opacity
\SetBgAngle{90.0}% Select rotation of logo
\SetBgScale{1.5}% Select scale factor of logo

\definechangesauthor[name={Ajay}, color=magenta]{Ajay}
\definechangesauthor[name={Neal}, color=blue]{Neal}
\definechangesauthor[name={KTB}, color=red]{KTB}
\definechangesauthor[name={Evelio}, color=green]{Evelio}
\definechangesauthor[name={Gilbert}, color=orange]{Gilbert}
\definechangesauthor[name={Jose}, color=purple]{Jose}

\newcommand{\Z}{\mathbb{Z}} % integers
\newcommand{\Zq}{\mathbb{Z}_{\q}} % finite field of size q
\newcommand{\N}{N} % size of coefficient vectors
\newcommand{\q}{q} % prime modulus
\newcommand{\floor}[1]{\lfloor{#1}\rfloor}
\newcommand{\ceil}[1]{\lceil{#1}\rceil}
\newcommand{\Mod}[1]{\ \mathrm{mod}\ #1}
\newcommand{\Modq}{\ \mathrm{mod}\ \q}
\newcommand{\numbits}[1]{\mathrm{len}(#1)} % number of bits
\newcommand{\wordsize}{\upbeta} % word size
\newcommand{\summ}{\mathrm{sum}}
\newcommand{\prodd}{\mathrm{rem}}
\newcommand{\quot}{\mathrm{quot}}
\newcommand{\barrettconstant}{\mu}
\newcommand{\ntt}[1]{\mathrm{NTT}_{#1}}

\newcommand{\butt}{\mathrm{butt}}
\newcommand{\buttct}[1]{\butt_{#1}^{\mathrm{CT}}}
\newcommand{\buttgs}[1]{\butt_{#1}^{\mathrm{GS}}}
\newcommand{\mnttct}[1]{\mathrm{NTT}^{\mathrm{CT},{#1}}_{no\rightarrow bo}}
\newcommand{\mnttgs}[1]{\mathrm{NTT}^{\mathrm{GS},{#1}}_{bo\rightarrow no}}
\newcommand{\snttgs}[1]{\mathrm{NTT}^{\mathrm{GS},{#1}}_{bo\rightarrow no,\frac{1}{2}}} % (1/N)-scaling merged NTT
\newcommand{\vect}{{\bf a}}

\newcommand{\vectimg}{{\bf \widehat{a}}}
\newcommand{\vecttwo}{{\bf b}}
\newcommand{\vectthree}{{\bf c}}
\newcommand{\vecttwoimg}{{\bf \widehat{b}}}
\newcommand{\vectthreeimg}{{\bf \widehat{c}}}
\newcommand{\roots}{{\bf \Psi}}
\newcommand{\br}{\mathrm{br}} % bit reversal

\newcommand{\rootsinv}{{\bf \Psi^{-1}}}
\newcommand{\tnttct}[1]{\mathrm{\widehat{NTT}}^{\mathrm{CT},{#1}}_{no\rightarrow bo}} % truncated ct ntt
\newcommand{\tnttgs}[1]{\mathrm{\widehat{NTT}}^{\mathrm{GS},{#1}}_{bo\rightarrow no,\frac{1}{2}}} % truncated gs ntt
\newcommand{\circconv}{\circledast}
\newcommand{\negaconv}{\circconv}

\begin{document}

\title{Accelerating Polynomial Multiplication for Homomorphic Encryption on GPUs}

\author{
\IEEEauthorblockN{
Kaustubh Shivdikar\IEEEauthorrefmark{1},
Gilbert Jonatan\IEEEauthorrefmark{2},
Evelio Mora\IEEEauthorrefmark{3},
Neal Livesay\IEEEauthorrefmark{1},
Rashmi Agrawal\IEEEauthorrefmark{4}, \\
Ajay Joshi\IEEEauthorrefmark{4},
Jos\'e L. Abell\'an\IEEEauthorrefmark{3},
John Kim\IEEEauthorrefmark{2},
David Kaeli\IEEEauthorrefmark{1}
}
\IEEEauthorblockA{
\IEEEauthorrefmark{1}Northeastern University,
\IEEEauthorrefmark{4}Boston University,
\IEEEauthorrefmark{2}KAIST University,
\IEEEauthorrefmark{3}Universidad Cat\'olica de Murcia \\
{\{shivdikar.k, n.livesay\}}@northeastern.edu,
{\{eamora, jlabellan\}}@ucam.edu, 
{\{rashmi23, joshi\}}@bu.edu, \\
kaeli@ece.neu.edu,
gilbertjonatan@kaist.ac.kr,
jjk12@kaist.edu
}
}

% \author{Name1}
% \email{email1@email.com}
% \affiliation{University name, Department name, Address}

% \author{Name2}
% \email{email2@email.com}
% \affiliation{University name3, Department name, Address}

% \author{Name3}
% \email{email3@email.com}
% \affiliation{University name2, Department name, Address}

\maketitle

%     _    _         _                  _   
%    / \  | |__  ___| |_ _ __ __ _  ___| |_ 
%   / _ \ | '_ \/ __| __| '__/ _` |/ __| __|
%  / ___ \| |_) \__ \ |_| | | (_| | (__| |_ 
% /_/   \_\_.__/|___/\__|_|  \__,_|\___|\__|

\begin{abstract}
%With the advent of quantum computing, large-scale quantum computers pose a threat to the reliability of current public-key cryptosystems, necessitating the need for quantum-proof cryptographic systems. Lattice-based cryptography is a promising framework for both post-quantum cryptography and Homomorphic Encryption (HE). Lattice-based FHE enables users to securely outsource both the storage and computation of sensitive data to untrusted servers. Although FHE offers an attractive solution for security in cloud systems, it suffers from prohibitively high computational costs. The primary bottleneck in lattice-based cryptography is polynomial multiplication. For lattice-based FHE to become viable for real-world systems, we need a highly efficient implementation of polynomial multiplication.

%Neal: here is an attempt to improve the flow of paragraph 
Homomorphic Encryption (HE) enables users to securely outsource both the storage and computation of sensitive data to untrusted servers. Not only does HE offer an attractive solution for security in cloud systems, but lattice-based HE systems are also believed to be resistant to attacks by quantum computers. However, current HE implementations suffer from prohibitively high latency. For lattice-based HE to become viable for real-world systems, it is necessary for the key bottlenecks---particularly polynomial multiplication---to be highly efficient.

%In this paper, we present a characterization of GPU-based implementations of polynomial multiplication targeting its two major computationally-expensive operations: modular reduction and the Number Theoretic Transform (NTT). We begin with a survey of modular reduction and analyze several variants of the widely-used Barrett modular reduction algorithm. Then, we propose a variant optimized for $62$-bit integer words on the GPU, obtaining a $1.22\times$ speedup over the existing comparable implementations. Next, we propose a set of incremental GPU-specific improvements of polynomial multiplication targeted at optimizing latency and throughput. We present a mixed-radix, 2D-NTT, multi-block implementation that results in a $1.36\times$ speedup for $N=2^{16}$ over the previous state-of-the-art. We explore shared memory optimizations aimed at reducing redundant memory accesses, further improving speedups by $1.25\times$.

In this paper, we present a characterization of GPU-based implementations of polynomial multiplication. We begin with a survey of modular reduction techniques and analyze several variants of the widely-used Barrett modular reduction algorithm. We then propose a modular reduction variant optimized for $64$-bit integer words on the GPU, obtaining a $1.8\times$ speedup over the existing comparable implementations. 
Next, we explore the following GPU-specific improvements for polynomial multiplication targeted at optimizing latency and throughput: 1) We present a 2D mixed-radix, multi-block implementation of NTT that results in a $1.85\times$ average speedup over the previous state-of-the-art. 2) We explore shared memory optimizations aimed at reducing redundant memory accesses, further improving speedups by $1.2\times$. 3) Finally, we fuse the Hadamard product with neighboring stages of the NTT, reducing the twiddle factor memory footprint by $50\%$. By combining our NTT optimizations, we achieve an overall speedup of $123.13\times$ and $2.37\times$ over the previous state-of-the-art CPU and GPU implementations of NTT kernels, respectively.
\end{abstract}


%  ___       _                 _            _   _             
% |_ _|_ __ | |_ _ __ ___   __| |_   _  ___| |_(_) ___  _ __  
%  | || '_ \| __| '__/ _ \ / _` | | | |/ __| __| |/ _ \| '_ \ 
%  | || | | | |_| | | (_) | (_| | |_| | (__| |_| | (_) | | | |
% |___|_| |_|\__|_|  \___/ \__,_|\__,_|\___|\__|_|\___/|_| |_|

\section{Introduction}
%The rate of new cyberattacks, which include phishing attacks, ransomware, and malware exploits, has been increasing steadily~\cite{branch2019trends}.
%The number of data breaches has increased by more than $190\%$ since 2020~\cite{itrc_2022}.
% The year 2021 marks the highest number of data breaches to date, with a staggering $293$ million records stolen in a single year~\cite{itrc_2022}.
%%The outsourcing of computation to servers is on the rise. Classical encryption methods protect data as it is transmitted over the internet. However, computation requires that data be decrypted on the (potentially untrusted) server.
%Current encryption schemes do not preserve client privacy as the data encrypted by the clients is decrypted by the cloud computing services.
%%Given growing security and privacy concerns in networked infrastructure, consumer trust in cloud computing services has been eroding~\cite{resende2021towards}.
%%The uncertainty associated with security of cloud computing can be a barrier to moving valuable corporate IP, sensitive computations, and customer information to public cloud-based services~\cite{5655238,jayaweera2021jaxed,thakkar2017video}.
% ~\cite{5655238,jayaweera2021jaxed, thakkar2017video}.
%Moreover, many of the breakthrough FHE schemes are lattice-based, and are believed to be resistant to attacks by quantum computers~\cite{cominetti2020fast}.

\begin{figure}[htbp]
	\centering
	\includegraphics[width=0.48\textwidth]{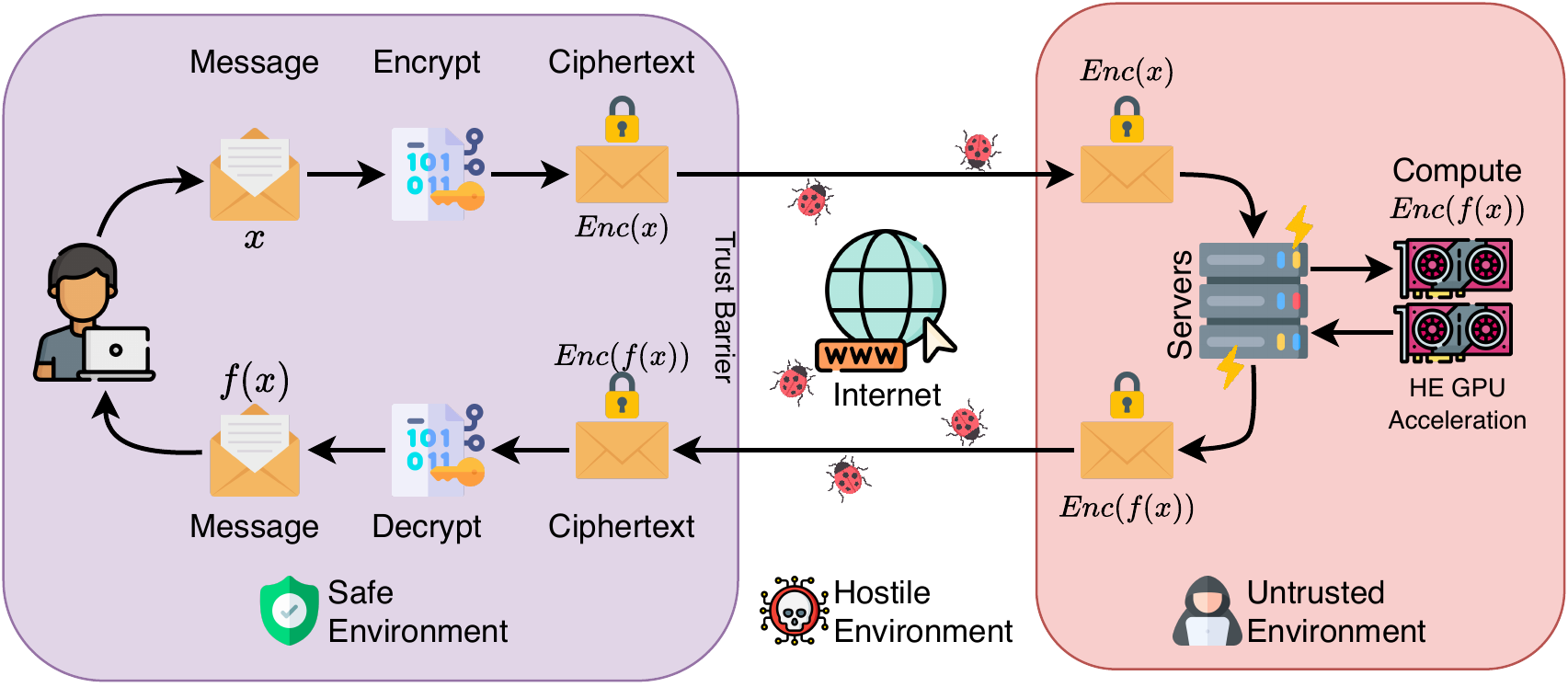}
	\vspace{-0.7em}
	\caption{HE provides security from eavesdroppers on the web as well as untrusted cloud services, as encrypted data can be computed on directly.}
	%\caption{FHE protects against network insecurities in untrusted cloud services, enabling users to securely offload sensitive data}
	\label{fig:fhe_overview}
\vspace{-1.9em}
\end{figure}

Computation is increasingly outsourced to remote cloud-computing services~\cite{5655238, branch2019trends}. Encryption provides security as data is transmitted over the internet. However, classical encryption schemes require that data be decrypted prior to performing computation, exposing sensitive data to untrusted cloud providers~\cite{jayaweera2021jaxed,thakkar2017video}.
Using Homomorphic Encryption (HE) allows computations to be run directly on encrypted operands, offering ideal security in the cloud-computing era~(Figure~\ref{fig:fhe_overview}). Moreover, many of the breakthrough HE schemes are lattice-based, and are believed to be resistant to attacks by quantum computers~\cite{cominetti2020fast}. 

%but require that data be decrypted prior to performing computation. This raises concerns while offloading sensitive computations to public cloud-based services
% However, classical schemes 
%Fully Homomorphic Encryption (FHE) can provide a path forward to protect applications in untrusted cloud environments~\cite{rahman2019privacy}. An FHE scheme allows computation directly on encrypted operands, without requiring them to first be decrypted. This offers ideal security in the cloud-computing era. Moreover, many of the breakthrough FHE schemes are lattice-based, and are believed to be resistant to attacks by quantum computers~\cite{cominetti2020fast}.

One major challenge in deploying HE in real-world systems is overcoming the high computational costs associated with HE. For computation on data encrypted via state-of-the-art HE schemes---such as HE for Arithmetic of Approximate Numbers~\cite{ckks} (also known as HEAAN or CKKS) and TFHE~\cite{chillotti2020tfhe}--- a slowdown of 4--6 orders of magnitude is reported, as compared to running the same computation on unencrypted data~\cite{f1,jung2021accel}. We aim to accelerate HE by targeting the main operation in these schemes (and, more generally, in lattice-based cryptography): \emph{polynomial multiplication}~\cite{gentry2009,lyubashevsky,longa}.
%Numerous FHE schemes have been proposed in recent years ~\cite{brakerski2014leveled, cheon2017homomorphic, chillotti2020tfhe, ducas2015fhew, fan2012somewhat}. HE for Arithmetic of Approximate Numbers (HEAAN~\cite{cheon2017homomorphic}), also known as Cheon--Kim--Kim--Song (CKKS Scheme) is gaining popularity since it supports approximate computation~\cite{kuo2021benchmarking, kuo2020idash, wang2018idash}. We aim to accelerate FHE by targeting the main operation in lattice-based FHE schemes (like HEAAN)---\emph{polynomial multiplication}~\cite{gentry2009,lyubashevsky,longa}. 
The Number Theoretic Transform (NTT) and modular reduction are two key bottlenecks in polynomial multiplication (and, by extension, in HE), as evidenced by the performance profiling of several lattice-based cryptographic algorithms by Koteshwara et al.~\cite{koteshwara}. As lattice-based HE schemes have continued to establish themselves as leading candidates for privacy-preserving computing and other applications, there has been an increased focus on optimization and acceleration of these core operations~\cite{darpa, kim2021general, kadykov2021homomorphic}.

%GPU platforms are natural candidates for accelerating FHE due to the inherent parallelism exhibited by FHE workloads.

%old version of para 3
% For most real-world applications of lattice-based FHE, the number $\N$ of polynomial coefficients, and the modulus $Q$ (which is directly related to the total workload size), need to be large to guarantee a strong level of security and sufficient compute levels~\cite{jung2021accel}. 
% For example, $\N=2^{16}$ and $\log(Q)=1240$ are the default values in the Homomorphic Encryption for Arithmetic of Approximate Numbers (HEAAN/CKKS) HE library.  These workload sizes require a large amount of computational power to evaluate modular arithmetic expressions, as well as high memory bandwidth to keep up with the data demands. Existing compute systems such as general-purpose CPUs do not scale well for these data-intensive workloads. However, the SIMD style GPU platforms, with their thousands of cores and high bandwidth memory (HBM), are natural candidates for accelerating these highly parallel workloads.

For most real-world applications of lattice-based HE, the number $\N$ of polynomial coefficients and the modulus $Q$ need to be large to guarantee a strong level of security and a higher degree of parallelism~\cite{jung2021accel}. 
For example, $\N=2^{16}$ and $\ceil{\log_2(Q)}=1240$ are the default values in the HEAAN library.
The large values for $N$ and $Q$ translate to heavy workload demands, requiring a significant amount of computational power to evaluate modular arithmetic expressions, as well as placing high demands on the memory bandwidth utilization. HE workloads possess high levels of data parallelism~\cite{martins2016enhancing}. Existing compute systems such as general-purpose CPUs do not scale well since they are unable to fully exploit this parallelism for such data-intensive workloads. However, the SIMD-style GPU platforms, with their thousands of cores and high bandwidth memory (HBM), are natural candidates for accelerating these highly parallelizable workloads.
The potential of the GPU platform to accelerate HE has motivated a rapidly growing body of work over the past year~\cite{jung2021accel,jung2021boot, zhai,kim2020,ozerk,durrani,sahu,goey,albadawi,lee}. 

\begin{figure}[t]
	\centering
	\vspace{-0.7em}
	\includegraphics[width=0.48\textwidth]{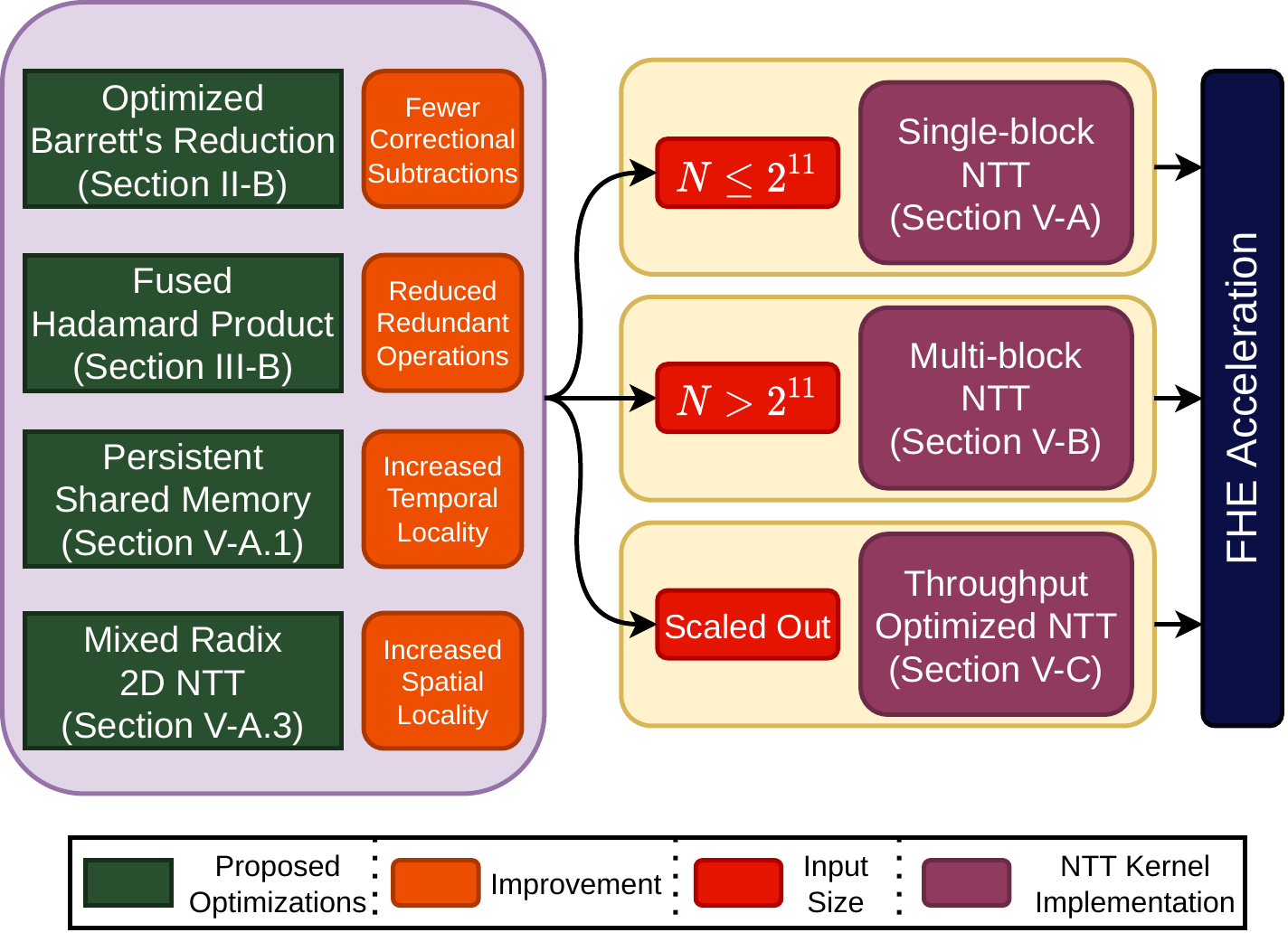}
	\vspace{-0.7em}
	\caption{Our contributions: $4$ major optimizations incorporated into $3$ kernels}
	\label{fig:contributions}
\vspace{-0.8em}
\end{figure}

To address performance bottlenecks in existing polynomial multiplication algorithms, we begin by analyzing the Barrett modular reduction algorithm~\cite{barrett}, as well as the algorithm's variants~\cite{ozerk,lee,dhem} which have been utilized in prior HE schemes.
% We then analyze various mixed-radix kernel implementations which focus on polynomial multiplication that we tuned for memory efficiency.
We then analyze various NTT implementations, including mixed-radix and 2D implementations, which we tune to improve memory efficiency.
%aim for high data locality.
% We explore recent algorithmic optimizations~\cite{??} and propose our own. 
% fusing polynomial multiplication with butterfly operations, providing a further reduction in the computational requirements of HE. 
Finally, we apply a number of GPU-specific optimizations to further accelerate HE. %Note that our implementations target a 64-bit word-size.
By combining all our optimizations, we achieve an overall speedup of $123.13\times$ and $2.37\times$ over the previous state-of-the-art CPU~\cite{sealcrypto} and GPU~\cite{ozerk} implementations of NTT kernels, respectively.
% We target the 32-bit wordsize.
%
%Kim et al.~\cite{kim2020} observed that the choice of whether to use 
%
%
% Much of the previous work has focused on FHE implementations for 64-bit words, perhaps due in part to an established trend for CPU.
% However, it seems far from settled whether 64-bit words are the optimal choice for GPU.
% Kim et al.~\cite{kim2020} note that using a 64-bit word in GPU implementations of \textcolor{blue}{NTT} (as opposed to a 32-bit word) offers a reduced workload size but at the cost of a higher operational complexity, which results in a ``negligible'' performance difference between the 32-bit and 64-bit implementations.
% \textcolor{blue}{In this paper, we focused on investigating relatively unexplored 32-bit word implementations and their performance on the GPU. The latest GPU architectures like the Volta and the Ampere architecture natively support 32 bit integer arithmetic instructions. The registers in these architecture are also $32$-bits wide. $64$-bit integer arithmetic operations on the GPU are emulated using a series of instructions rather than being natively supported as their $32$-bit counter-parts.} 
% \comment[id=Neal]{is this last sentence awkward? Also: Perhaps Kaustubh should say some hardware stuff in here.}
%
%We analyze the performance of the GPU implementations of polynomial multiplication and its key component functions: modular reduction and the Number Theoretic Transform. 
Our key contributions are as follows (Figure~\ref{fig:contributions}):
\begin{enumerate} 
\item  We propose an instantiation of the Dhem--Quisquater~\cite{dhem} class of Barrett reduction variants which is optimized for HE,
%, especially for 64-bit word sizes.
providing a $1.85\times$ speedup over prior studies~\cite{ozerk,sahu,lee,wang}.
%for moduli of size up to 30 bits, the moduli size used in prior studies~\cite{ozerk,sahu,lee}. Our proposed Barrett reduction variant achieves a $1.22\times$ speedup over 
%the next fastest general modulus  Barrett reduction algorithm. \comment[id=Neal]{Give some numbers for polynomial mult with our method vs classical}
%We experiment with radix $2$, $4$, $8$, and $16$ implementations as %well as hybrid combinations of radix $4-16$ and radix $8-16$. 
%These higher radix kernels exhibit high temporal locality in their data access patterns. 
%Our best performing kernel of hybrid radix $8-16$  outperforms the other implementations, with a 
%We present a mixed-radix implementation to maximize temporal and spatial locality for efficient L1 cache utilization leading to a $1.36\times$ speedup over the radix-2 baseline.
\item We present a mixed-radix, 2D NTT implementation to effectively exploit temporal and spatial locality, resulting in a $1.91\times$ speedup over the radix-2 baseline.
%classical radix-$2$ implementation.
%% Higher radix implementations impact on the H/W
\item We propose a \emph{fused polynomial multiplication} algorithm, which \emph{fuses} the Hadamard product with its neighboring butterfly operations using an application of Karatsuba's Algorithm~\cite{karatsuba}. This reduces the twiddle factor's memory footprint size by $50\%$.
\item We incorporate the use of low latency, persistent, shared memory in our single-block NTT GPU kernel implementation, reducing the number of redundant data fetches from global memory, providing a further $1.25\times$ speedup.
\end{enumerate}

\section{Barrett reduction and its variants}
\label{sec:barrettsec}

%Neal: next part rewritten
Modular reduction is a key operation and computational bottleneck in lattice-based cryptography~\cite{cheon2018bootstrapping}. This section is a self-contained survey of modular reduction algorithms, particularly Barrett reduction~\cite{barrett}, a widely-used algorithm that we utilize in our work.
%which is widely used in FHE applications and utilized in our work.

Following Shoup~\cite{shoupbook}, we define the \emph{bit length} $\numbits{a}$ of a positive integer $a$ to be the number of bits in the binary representation of $a$; more precisely, $\numbits{a}=\floor{\log_2a}+1$.

\subsection{Background: modular reduction and arithmetic}
Let $x\Modq$ denote the remainder of a nonnegative integer $x$ divided by a positive integer $\q$.
The naive method for performing \emph{modular reduction}---i.e., the computation of $x\Modq$---is via an integer division operation: 
\[x\Modq =x-\floor{x/\q}\q.\]
However, there are a number of alternative methods for performing modular reduction---especially in conjunction with arithmetic operations such as addition and multiplication---that avoid expensive integer division operations. %Different methods may be preferable depending on the algorithmic context and hardware platform. 

%Note that, if $a,b>0$, then $\max\{\numbits{a},\numbits{b}\}\le\numbits{a+b}\le\max\{\numbits{a},\numbits{b}\}+1$ and $\numbits{a\times b}=\numbits{a}+\numbits{b}$. Thus some care is required in implementing modular reduction with arithmetic to avoid integer overflow. In this paper, we assume a conservative approach to modular arithmetic by imposing the condition that the operands and outputs for each operation are reduced (i.e., in $[0,q)$), rather than performing multiple operations before reducing. Note that it may be possible to modify a conservative implementation to make it ``lazy'' for significant improvements in computational efficiency; see, e.g., \cite{harvey2014}.

For example, Algorithm~\ref{alg:add} specifies a simple and efficient computation of the modular reduction of a sum.  Let $\wordsize$ denote the word-size (e.g., $\wordsize=32$ or $64$). 
%(we assume $\upbeta=32$ on the GPU). 
% The restriction $\numbits{\q}\le\wordsize-1$ prevents overflow of the intermediate operations (i.e., $a+b$).
% In our experiments, we found that Algorithm~\ref{alg:add} can provide $1.13\times$ speedup over the naive modular addition (i.e., $(a+b)\%\q$) for 30-bit moduli $\q$.
Observe that either $a+b$ lies in $[0,\q)$ and is reduced, or $a+b$ lies in $[\q,2\q)$ and requires a single \emph{correctional subtraction} to become reduced (see lines 2--3). 
The restriction $\numbits{\q}\le\wordsize-1$ prevents overflow of the transient operations (i.e., $a+b$).
%Note that conditional additions/subtractions can be implemented without branching (e.g., lines 2--3 can be replaced by $\summ \leftarrow \summ - (\summ\ge\q)\times\q$), provided that the conditional expression $(\summ\ge\q)$ evaluates to $1$ if true and $0$ otherwise. 
%\rashmi{May be add a sentence about why branching is to be avoided?}

\begin{algorithm}
\caption{A baseline modular addition algorithm}\label{alg:add}
\begin{algorithmic}[1]
\REQUIRE $0\le a,b<\q$, $\numbits{\q}\le \wordsize - 1$
\ENSURE $\summ = (a+b)\Modq$
\STATE $\summ \gets a+b$
\IF{$\summ\ge\q$}
    \STATE $\summ \gets \summ - \q$
\ENDIF
\RETURN $\summ$
\end{algorithmic}
\end{algorithm}

There are multiple methods for reducing products. In lattice-based cryptography, commonly used algorithms for implementations on hardware platforms such as CPUs and GPUs include the algorithms of Barrett~\cite{barrett}, Montgomery~\cite{montgomery}, and Shoup~\cite{shoup, harvey2014}.
In this paper, we select Barrett's algorithm as our baseline,
%modular reduction for our implementation of polynomial multiplication
 as Barrett's algorithm enjoys the following features:
\begin{enumerate}
\item \emph{Low overhead:} It requires a low-cost pre-computation (and storage) of a single word-size integer $\barrettconstant$.
\item \emph{Versatility:} It may be used effectively in contexts where multiple products
%---or, more generally, integers in $[0,2^{2\numbits{\q}})$---
are reduced modulo $\q$.
\item \emph{Generality:} 
%Some modular reduction algorithms allow only specialized moduli, such as Mersenne primes (e.g., \cite{acar,knezevic}). 
It does not restrict to special classes of moduli, such as Mersenne primes (see, e.g., \cite{acar,knezevic}).
%\item \emph{Ease-of-use:} Barrett reduction function can be easily called with the pre-computed constant $\barrettconstant$ to calculate $x \mod\q$.
\item \emph{Performant:} It is significantly faster than integer division, and has comparable runtime performance with Montgomery's algorithm (see, e.g., \cite{hars}).
\end{enumerate}
Barrett's algorithm is used in many open-source libraries, including cuHE~\cite{cuhe}, PALISADE~\cite{palisade}, and HEANN~\cite{kim2020, ckks}. The Barrett reduction algorithm, and our proposed variant for use in HE, are analyzed in Section~\ref{sec:barrett}.

% introductory sentence
% TODO: change this section title
\subsection{Barrett modular reduction: analysis and optimization}
\label{sec:barrett} 

\begin{figure*}[htbp]
	\centering
\includegraphics[width=1.0\textwidth]{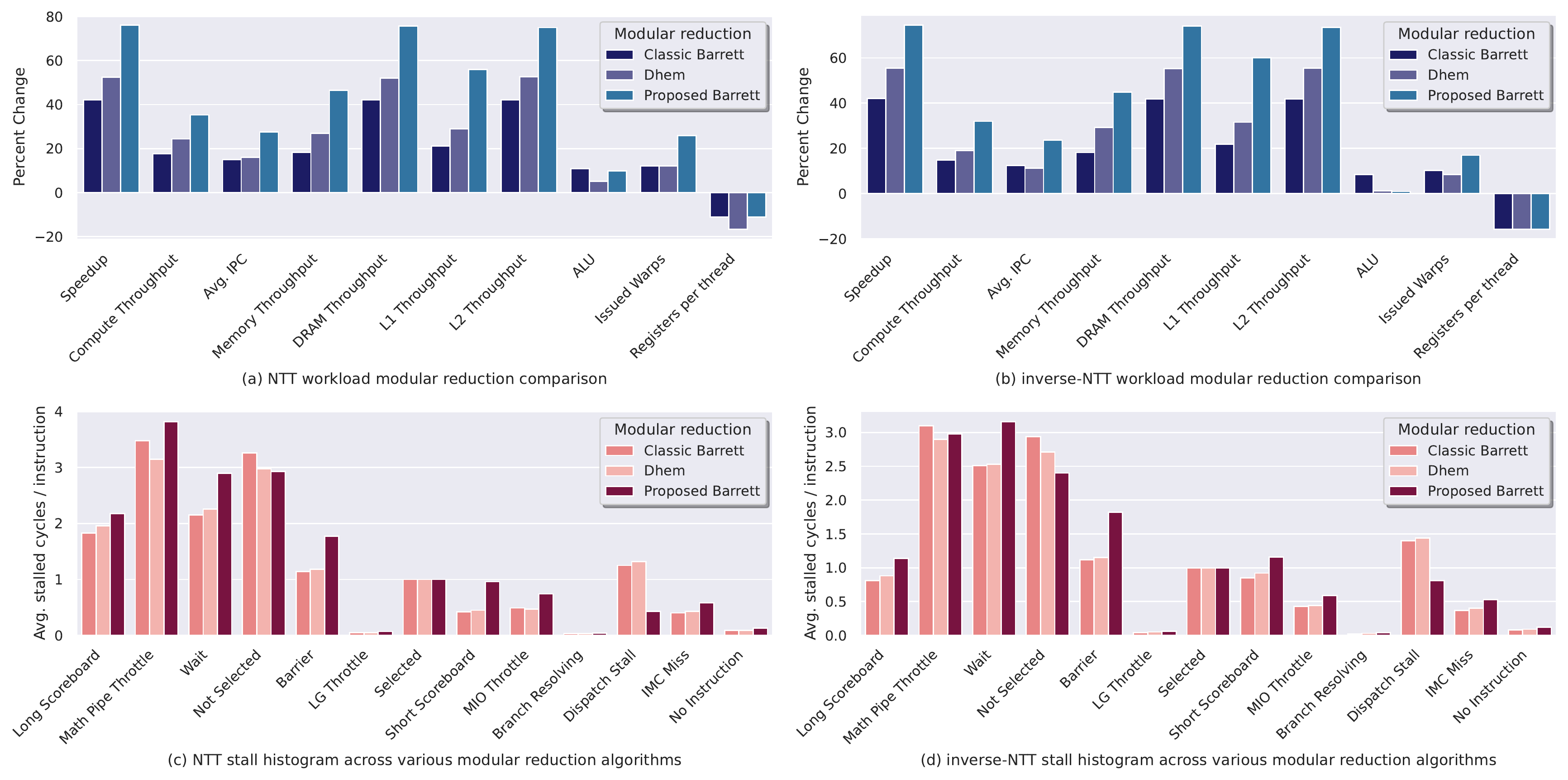}
	\vspace{-2em}
	\caption{Modular reduction profile comparison of architectural parameters (a,b) and causes of warp stalls (c,d).}
	\label{fig:mod_red_all}
\vspace{-1.6em}
\end{figure*}

Next, we provide details of Barrett modular reduction and then explore potential improvements.
Algorithm~\ref{alg:barrett} specifies the classical reduction algorithm of Barrett~\cite{barrett}. 
%Barrett reduction (with respect to a modulus $\q$) requires a one-time pre-processing step: the computation (and storage) of a word-sized integer $\barrettconstant$. 
%As described in Algorithm~\ref{alg:barrett},
\begin{algorithm}
\caption{Classical Barrett reduction}\label{alg:barrett}
\begin{algorithmic}[1]
\REQUIRE $m=\numbits{q}\le\wordsize -2$, $0\le x<2^{2m}$,  $\barrettconstant=\floor{\frac{2^{2m}}{\q}}$
\ENSURE $\prodd = x\Modq$
\STATE $c \gets x\gg(m-1)$
\STATE $\quot \gets (c\times\barrettconstant)\gg(m+1)$
\STATE $\prodd \gets x - \quot\times\q$
\IF{$\prodd\ge\q$}
    \STATE $\prodd \gets \prodd - \q$
\ENDIF
\IF{$\prodd\ge\q$}
    \STATE $\prodd \gets \prodd - \q$
\ENDIF
\RETURN $\prodd$
\end{algorithmic}
\end{algorithm}
Note that if $0\le a,b<\q$ and $m=\numbits{\q}$, then $x=a\times b$ satisfies the condition $0\le x<2^{2m}$ specified in Algorithm~\ref{alg:barrett}. This algorithm is commonly used in HE acceleration studies targeting a GPU~\cite{goey,wang,sahu,dai2014}.
As noted by Sahu et al.~\cite{sahu}, the pre-computed constant $\mu$ and the transient operations (excluding the product $c\times \barrettconstant$) are preferably word-sized. This condition imposes the restriction $\numbits{\q}\le\wordsize-2$.

% \begin{table}[b]
% \vspace{-2em}
% \centering
% \caption{Correctional subtractions required for modular reduction.}
% \begin{tabular}{c c c c}
% \hline\hline %inserts double horizontal lines
%   Number of & Classical & \"{O}zerk & Proposed \\
%   correctional & Barrett & et al. & Barrett \\
%   subtractions & (Algo. \ref{alg:barrett}) & \cite{ozerk,ozerkcode} & (Algo. \ref{alg:barrettfast}) \\
% \hline % inserts single horizontal line

% 0 & 41.5\% & 58.7\% & 70.1\% \\ [1ex]
% 1 & 56.5\% & 41.1\% & 29.9\% \\ [1ex]
% 2 & 2.1\% & 0.21\% & 0\% \\ [1ex]
% \hline % inserts single horizontal line
% \vspace{1mm}
% \end{tabular}
% \label{tab:subtractions}
% \end{table}

Note that the classical Barrett reduction may require zero, one, or two correctional subtractions; see lines 4--7 in Algorithm~\ref{alg:barrett}.
%The second conditional subtraction specified in lines 6--7 is performed in approximately 1\% of the cases~\cite{barrett, omondi}.
%The classical Barrett reduction algorithm can be viewed as ``almost'' not requiring the second conditional subtraction specified in lines 6--7, as Barrett~\cite{barrett} reports that a second conditional subtraction is performed in approximately $1\%$ of cases (see also Omondi et al.~\cite{omondi}).
%In Section~\ref{sec:polymult}, we restrict our attention to specialized moduli $\q$ that are suitable for efficiently computing polynomial multiplication (specifically, moduli $\q$ with the property that $\q$ is prime and $2\times2^{16}$ is a divisor of $\q-1$).
% We performed an experiment (where the input and moduli were each randomly sampled from uniform distributions in 10 billion trials) to approximate the probability that a second correctional subtraction is required for $30$-bit moduli.  Results of our experiment are provided in Table~\ref{tab:subtractions}. We found that the classical Barrett reduction requires two conditional subtractions to fully reduce the input in $2\%$ of the cases. %Barrett~\cite{barrett} and Omondi~\cite{omondi} both report that it is possible to prove mathematically 
As noted by Barrett~\cite{barrett}, a second conditional subtraction is required in approximately 1\% of the cases. There have been several attempts to modify Barrett's algorithm to eliminate the need for a second conditional subtraction.
%\"{O}zerk et al.~\cite{ozerk} propose a variant (see also their open-source code~\cite{ozerkcode}) that nearly achieves at most one correctional subtraction, but we found experimentally that their implementation requires two correctional subtractions in $0.22\%$ of cases. 
The algorithms proposed by \"{O}zerk et al.~\cite{ozerk} and Lee et al.~\cite{lee} each require two correctional subtractions to fully reduce the product of $a=994674970$ and $b=994705408$ modulo $q=994705409$, although we found experimentally that \"{O}zerk et al.'s proposed reduction algorithm only requires a second conditional subtraction in $0.22\%$ of cases.

%% Barrett Reduction comparisons
% \begin{figure}[htbp]
% 	\centering
% 	\includegraphics[width=0.48\textwidth]{figures/03_barrett/CPU_plot.pdf}
% 	\caption{Execution times for various modular reduction implementations on CPU}
% 	\label{fig:modular_mult_cpu}
% \end{figure}

Dhem--Quisquater~\cite{dhem} defines a class of Barrett modular reduction variants (with parameters $\alpha$ and $\beta$) that require at most one correctional subtraction. A commonly used (see, e.g., Kong and Philips~\cite{kong} and Wu et al.~\cite{wu}) instantiation of Dhem--Quisquater's class of algorithms is specified in Algorithm \ref{alg:dhem} (setting parameters $\alpha=\N+3$ and $\beta=-2$, as defined in Dhem--Quisquater~\cite{dhem}). 
Notably, this instantiation is used in the PALISADE HE Software Library~\cite{palisade}. 
\begin{algorithm}[t]
\caption{Dhem–Quisquater's modified Barrett reduction}\label{alg:dhem}
\begin{algorithmic}[1]
%% ********* Do not remove color coding in equation. It is part of the paper. It signifies difference compared to Algo 3.
\REQUIRE $m=\numbits{q}\le\wordsize -4$, $0\le x<2^{2m}$,  $\barrettconstant=\floor{\frac{2^{2m+3}}{\q}}$
\ENSURE $\prodd = x\Modq$
\STATE $c \gets x\gg (m-2)$
\STATE $\quot \gets (c\times\barrettconstant)\gg(m+5)$
\STATE $\prodd \gets x - \quot\times\q$
\IF{$\prodd\ge\q$}
    \STATE $\prodd \gets \prodd - \q$
\ENDIF
\RETURN $\prodd$
\end{algorithmic}
\end{algorithm}
%In our experiments, Algorithm~\ref{alg:dhem} has a $1.22\times$ speedup over Algorithm~\ref{alg:barrett} for $28$-bit moduli..
Although Algorithm~\ref{alg:dhem} provides an improvement in algorithmic complexity over Algorithm~\ref{alg:barrett}, it further restricts the modulus to at most length $(\wordsize-4)$ to ensure $\barrettconstant$ is word-sized.

As discussed by Kim et al.~\cite{kim2020}, restrictions on the modulus size are significant in the context of optimizing HE, as the modulus size is inversely related to the \emph{workload size}. To elaborate, polynomial multiplication is typically performed with respect to a large composite modulus $Q$. If each prime factor of $Q$ is $m$-bits, then the Chinese Remainder Theorem can be used to partition the computation of polynomial multiplication with respect to $Q$ into $\ceil{\numbits{Q}/m}$ simpler computations of polynomial multiplication with respect to the $m$-bit factors.
For example, if $\numbits{Q}=1240$, then the restriction from 30-bit to 28-bit moduli increases the workload size (i.e., $\ceil{\numbits{Q}/m}$) by $7.14\%$.

%when $m$ is the modulus size. Thus it is prudent to consider both operational complexity and workload size when choosing a modular reduction algorithm for use in HE. Although Algorithm~\ref{alg:dhem} has a $1.22\times$ speedup over Algorithm~\ref{alg:barrett} for $28$-bit moduli, the restriction from 31-bit to 28-bit moduli increases the workload size by $10\%$.

Therefore, we propose Algorithm~\ref{alg:barrettfast} for use in HE implementations on a GPU. Similar to Algorithm~\ref{alg:dhem}, Algorithm~\ref{alg:barrettfast} is an instantiation of Dhem--Quisquater~\cite{dhem} (for $\alpha=\N+1$ and $\beta=-2$) that requires at most one correctional subtraction. However, Algorithm~\ref{alg:barrettfast} allows for moduli $\q$ of length up to $\wordsize-2$, and thus results in no increase in the workload size.
%The restriction from 31-bit to 30-bit moduli required by Algorithm~\ref{alg:barrettfast} increases the workload size by only $5\%$, as opposed to the $12.5\%$ increase imposed by Algorithm~\ref{alg:dhem}.

% to allow for moduli $\q$ of at most a $(\wordsize-2)$-bit (i.e., 30-bit) integer. 
% Algorithm \ref{alg:barrettfast} is an instantiation of Dhem--Quisquater \cite{dhem} (for parameters $\alpha=\N+1$ and $\beta=-2$). 

\begin{algorithm}[t]
\caption{Proposed Barrett reduction optimized for a GPU}\label{alg:barrettfast}
\begin{algorithmic}[1]
%% ********* Do not remove color coding in equation. It is part of the paper. It signifies difference compared to Algo 3.
\REQUIRE $m=\numbits{q}\le\wordsize-2$, $0\le x<2^{2m}$,  $\barrettconstant=\floor{\frac{2^{2m+1}}{\q}}$
\ENSURE $\prodd = x\Modq$
\STATE $c \gets x\gg (m-2)$ %TODO: make it clear that m-2 and m+3 are precomputed?
\STATE $\quot \gets (c\times\barrettconstant)\gg (m+3)$
\STATE $\prodd \gets x - \quot\times\q$
\IF{$\prodd\ge\q$}
    \STATE $\prodd \gets \prodd - \q$
\ENDIF
\RETURN $\prodd$
\end{algorithmic}
\end{algorithm}
%To prove correctness of Algorithm~\ref{alg:barrettfast}, it suffices to show that the ``approximate quotient'' $\quot$ computed in line 2 is sufficiently close to the true quotient $\floor{x/q}$; i.e., it suffices to prove $\floor{x/q}\ge\quot\ge\floor{x/q}-1$. Observe that
% \[\begin{array}{rll}
%  \floor{x/q} &= \text{ }\left\lfloor\frac{\frac{x}{2^{m-2}}\times\frac{2^{2m+1}}{\q}}{2^{m+3}}\right\rfloor \\
%     &\ge\left\lfloor\frac{\floor{\frac{x}{2^{m-2}}}\floor{\frac{2^{2m+1}}{\q}}}{2^{m+3}}\right\rfloor = \quot \\ 
%     & \ge \frac{1}{2^{m+3}}\left(\frac{x}{2^{m-2}}-1\right)\left(\frac{2^{2m+1}}{\q}-1\right)-1 \\
%     & = x/\q - \frac{x}{2^{2m+1}} - \frac{2^{m-2}}{\q} + \frac{1}{2^{m+3}} - 1 \\
%     & > \floor{x/\q} - 2.
% \end{array}\]
% Since $\quot$ is an integer, the claim follows.

Figure~\ref{fig:mod_red_all} provides a snapshot of the performance of various modular reduction kernels on a V100 GPU. The detailed description of each parameter is further described in Table~\ref{table:arch_profile} and \ref{table:stall_profile} in Section~\ref{sec:gpu}.
The values in the Figure~\ref{fig:mod_red_all}(a,b) are normalized to the built-in implementation of modular reduction on GPUs (which utilizes the modulo ``$\%$'' operator).
In Figure~\ref{fig:mod_red_all}(a,b) we see significant improvements in the proposed Barrett reduction, as marked by the speedups due to improved compute and memory throughput. The performance improvements achieved
%is a combined contribution of increased memory and compute throughput which 
can be attributed to our implementation requiring at most $1$ correctional subtraction (as compared to $2$ for others).
Figure~\ref{fig:mod_red_all}(c,d) enable us to see the primary causes of kernel stalls for NTT and inverse-NTT workloads, respectively. Figure~\ref{fig:mod_red_all}(c,d) highlights the reasons for the maximum number of stalls while executing NTT and inverse-NTT kernels.
In Figure~\ref{fig:mod_red_all}(c), the longest stall (measured in the average number of cycles per instruction) for the NTT workload is due to a ``Math Pipe Throttle'', which results when the kernel begins to saturate the ALU instruction pipeline (See Table~\ref{table:arch_profile}).
Figure~\ref{fig:mod_red_all}(d) reports the cause of stalls in inverse-NTT, with the longest stall caused by a ``Wait'', which signifies the scheduler has an abundance of ``Ready'' warps and is starting to saturate the streaming multiprocessors (SMs) (See Table~\ref{table:stall_profile}).

\begin{figure}[t]
	\centering
	\includegraphics[width=0.5\textwidth]{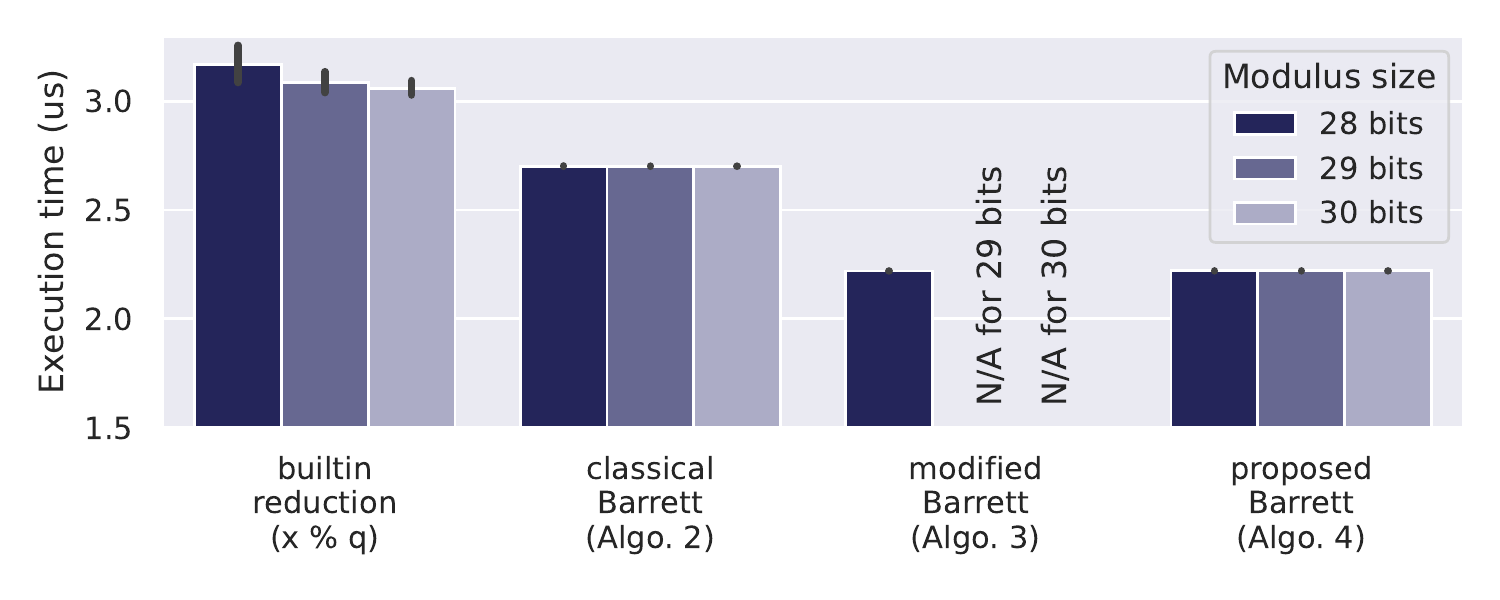}
	\vspace{-2.8em}
	\caption{Execution times of modular reduction implementations for $28$, $29$, and $30$-bit prime numbers (on the V100 GPU), averaged over 10,000 iterations. The error bars represent ranges. The ``builtin reduction'' uses the CUDA \% construct for modular reduction.}
	\label{fig:modular_mult_gpu}
\vspace{-1.5em}
\end{figure}

% \neal{TODO: update this section with numbers for 64-bit word}
In Figure~\ref{fig:modular_mult_gpu}, we present a comparison of the implementations of the modular reduction algorithms described in this section. We report the execution time of a single modular reduction operation for 28, 29, and 30-bit prime numbers as run on a V100 GPU. The operands and moduli are randomly sampled from a uniform distribution.
The classical Barrett reduction algorithm is significantly faster than reduction by integer division (i.e., the built-in reduction), as shown in Figure~\ref{fig:modular_mult_gpu}. 
Algorithm~\ref{alg:barrettfast} has nearly identical performance to Algorithm~\ref{alg:dhem} for 28-bit moduli (while permitting 29 and 30-bit moduli, as well).
%We found the difference in execution time between Algorithm~\ref{alg:barrettfast} and Algorithm~\ref{alg:dhem} to be negligible for 60-bit moduli. 
Algorithm~\ref{alg:barrettfast} has a $1.22\times$ speedup over the classical Barrett reduction for 30-bit primes.
%, while the restriction from 31- to 30-bit moduli only increases the workload size by $2.6\%$ (assuming $\numbits{Q}=1240$ as above).
To our knowledge, the specific instantiation of Dhem--Quisquater modular reduction specified in Algorithm~\ref{alg:barrettfast} does not appear in an open-source library nor in the literature.
%(other than as an instantiation of the class of Barrett variants defined by Dhem--Quisquater~\cite{dhem})

%and is the fastest Barrett modular reduction for general $30$ or $62$-bit moduli.
%In particular, Algorithm~\ref{alg:barrettfast} is faster than the Barrett reductions presented in prior work~\cite{sahu,ozerk,lee,goey}.
%%%%%%% 
%\neal{TODO: move setup and discussion for Figure 1 here}
%In our experiments, Algorithm~\ref{alg:dhem} has a $1.22\times$ speedup over Algorithm~\ref{alg:barrett} for $28$-bit moduli.

%  ____       _                             _       _ 
% |  _ \ ___ | |_   _ _ __   ___  _ __ ___ (_) __ _| |
% | |_) / _ \| | | | | '_ \ / _ \| '_ ` _ \| |/ _` | |
% |  __/ (_) | | |_| | | | | (_) | | | | | | | (_| | |
% |_|   \___/|_|\__, |_| |_|\___/|_| |_| |_|_|\__,_|_|
%               |___/                                 
%  __  __       _ _   _       _ _           _   _             
% |  \/  |_   _| | |_(_)_ __ | (_) ___ __ _| |_(_) ___  _ __  
% | |\/| | | | | | __| | '_ \| | |/ __/ _` | __| |/ _ \| '_ \ 
% | |  | | |_| | | |_| | |_) | | | (_| (_| | |_| | (_) | | | |
% |_|  |_|\__,_|_|\__|_| .__/|_|_|\___\__,_|\__|_|\___/|_| |_|
%                      |_|               

\section{Polynomial Multiplication}\label{sec:polymult}
For $m>0$, define $\Z_m$ to be the set $\{0,1,2,\ldots,m-1\}$ together with the operations of modular addition $(a,b)\mapsto (a+b)\Mod m$ and modular multiplication $(a,b)\mapsto (a\times b)\Mod m$.
The naive algorithm for multiplying
%(or \emph{convolving})
two polynomials $\sum_{i=0}^{\N-1}a_ix^i$ and $\sum_{i=0}^{\N-1}b_ix^i$
%(with coefficients $a_i$ and $b_i$ belonging to the set of, e.g., the real numbers or $\Z_m$),
 requires order $\N^2$ arithmetic operations.
It is well known~\cite{clrs} that the number of operations can be reduced to the order of $\N\log(\N)$ using the Fast Fourier Transform (FFT) algorithm. 
%To perform an FFT, the set of coefficients must contain a \emph{primitive $\N$th root of unity}. A primitive $\N$th root of unity in $\Z_m$ is an element $\omega$ satisfying the following properties:
%\begin{enumerate}
%\item $\omega^{\N}\Mod m = 1$; and 
%\item $\omega^i\Mod m\neq\omega^j\Mod m$ for all $0\le i<j<\N$.
%\end{enumerate}

It is convenient to represent a polynomial $\sum_{i=0}^{\N-1}a_ix^i$ as an $\N$-dimensional \emph{coefficient vector} $\vect=(a_0,a_1,\ldots,a_{\N-1})$.

%Let $\Zq[x]$ be the polynomials with coefficients in $\Zq$. Any element of the set $\Zq[x]/\langle x^{\N}-1\rangle$ of polynomials modulo $x^{\N}-1$ can be represented by an expression of the form $\sum_{i=0}^{\N-1}a_ix^i$ for some coefficients $a_i$ in $\Zq$. Thus, there is a one-to-one correspondence between $\Zq[x]/\langle x^{\N}-1\rangle$ and $(\Zq)^{\N}$ given by identifying the polynomial $\sum_{i=0}^{\N-1}a_ix^i$ with its \emph{coefficient vector} $\vect=(a_0,a_1,\ldots,a_{\N-1})$ in $(\Zq)^{\N}$. Likewise, the set \[\Rq=\Zq[x]/\langle x^{\N}+1\rangle\] can be identified with $(\Zq)^{\N}$ in the same way. The product in $\Zq[x]/\langle x^{\N}-1\rangle$ is called \emph{cyclic convolution}, and the product in $\Rq$ is called \emph{negacyclic convolution}~\cite{vonzurgathen}. We use the term \emph{polynomial multiplication} to mean negacyclic convolution, following the convention used in lattice-based cryptography~\cite{poppelmann}.

\begin{figure*}[htbp]
	\centering
	\includegraphics[width=1.0\textwidth]{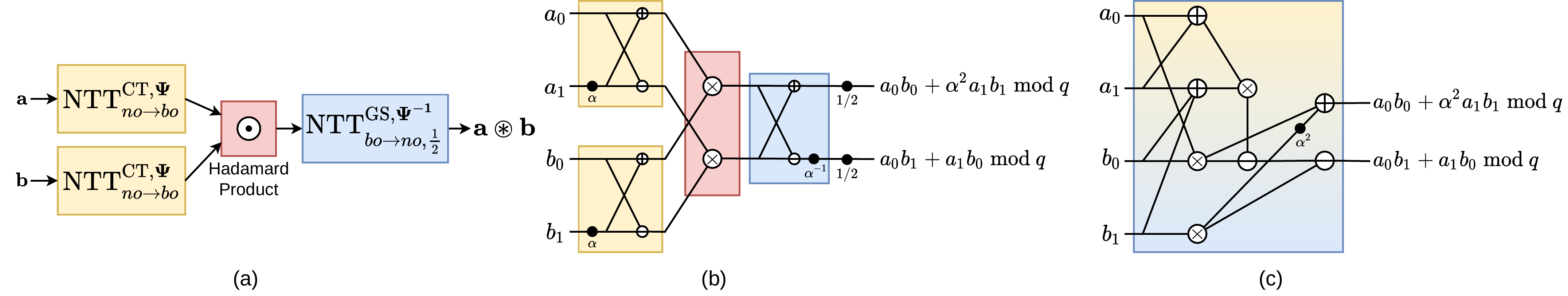}
	\caption{(a) Negacyclic convolution block diagram. (b) Hadamard product and its neighboring butterflies. (c) Fusion of butterflies into Hadamard product.}
	\label{fig:fused_ntt}
\vspace{-1.0em}
\end{figure*}

\subsection{Background: Number Theoretic Transform}
In this section, we give a brief review of the Discrete Fourier Transform (DFT) and the Fast Fourier Transform (FFT) for the special case that the field of coefficients is $\Zq$, for $\q$ a prime. The DFT and FFT over $\Zq$ are both commonly---and often confusingly---referred to as the Number Theoretic Transform (NTT). In the classical setup for the NTT, the parameters $\N$, $\q$, and $\omega$ satisfy the following properties:
\begin{enumerate}
\item $\N>1$ is a power of $2$;
\item $\q$ is a prime number such that $\N$ divides $\q-1$; and
\item $\omega$ is a \emph{primitive $\N$th root of unity} in $\Zq$; i.e., $\omega^i=1$ if and only if $i$ is a multiple of $\N$.
\end{enumerate}
The \emph{$\N$-point NTT (DFT) with respect to $\omega$} is the function $\ntt{\omega}:(\Zq)^{\N}\rightarrow(\Zq)^{\N}$ defined by $\ntt{\omega}(\vect)=(\sum_{i=0}^{\N-1}\vect[i]\omega^{ij})_{j=0}^{\N-1}$.  %$\vectimg=\ntt{\omega}(\vect)$ defined by $\vectimg[i]=\sum_{i=0}^{\N-1}\vect[j]\omega^{ij}\Modq$ for all $i$ in $[0,\N)$. 
The inverse transformation of $\ntt{\omega}$ is  $\frac{1}{\N}\ntt{\omega^{-1}}$. 
%Hence, $\ntt{\omega}\circ\frac{1}{\N}\ntt{\omega^{-1}} = \frac{1}{\N}\ntt{\omega^{-1}}\circ\ntt{\omega} = \Id$, where $\circ$ denotes composition of functions and $\Id$ is the identity. 
Famously, the cyclic convolution~\cite{vonzurgathen} of vectors $\vect$ and $\vecttwo$ in $(\Zq)^{\N}$ can be computed in the order of $\N\log(\N)$ arithmetic operations via the expression $\frac{1}{\N}\ntt{\omega^{-1}}(\ntt{\omega}(\vect)\odot\ntt{\omega}(\vecttwo))$, where $\odot$ denotes the Hadamard product (i.e., entry-wise multiplication) on $(\Zq)^{\N}$. 
% note: it is easy to forget the mult is modulo q

A closely related operation to cyclic convolution is \emph{negacyclic convolution}, which is widely known as \emph{polynomial multiplication} in the context of lattice-based cryptography~\cite{poppelmann}. 
%P{\"o}ppelmann et al. refers to as the {\em main operation}~\cite{poppelmann} in lattice-based cryptography. 
The setup for polynomial multiplication has parameters $\N$, $\q$, and $\psi$ satisfying the following properties:
\begin{enumerate}
\item $\N>1$ is a power of $2$;
\item $\q$ is a prime such that $2\N$ is a divisor of $\q-1$; and
\item $\psi$ is a primitive $2\N$th root of unity in $\Zq$ (which implies that $\omega=\psi^2$ is a primitive $\N$th root of unity).
\end{enumerate}
%To define the analogous relation between the NTT and polynomial multiplication (i.e., negacyclic convolution), suppose that $2\N$ divides $\q-1$. Then $\Zq$ has a primitive $2\N$th root of unity $\psi$ (as well as a primitive $\N$th root of unity $\omega=\psi^2$). 
Let $\roots$ and $\rootsinv$ denote the vector of ``twiddle factors'' in $(\Zq)^{\N}$, defined by $\roots[i]=\psi^i$ and $\rootsinv[i]=\psi^{-i}$ for all $i$. Then the negacyclic convolution $\vect\negaconv\vecttwo$ of vectors $\vect$ and $\vecttwo$ in $(\Zq)^{\N}$ satisfies the following relation \cite{crandall}: 
\begin{equation*}\vect\negaconv\vecttwo = \rootsinv\odot\frac{1}{\N}\ntt{\omega^{-1}}(\ntt{\omega}(\roots\odot\vect)\odot\ntt{\omega}(\roots\odot\vecttwo))\end{equation*}
%Henceforth, we refer to negacylic convolution as \emph{polynomial multiplication}, following a prevalent convention in the lattice-based cryptography literature~\cite{poppelmann}.

\begin{figure}[htbp]
	\centering
	\includegraphics[width=0.45\textwidth]{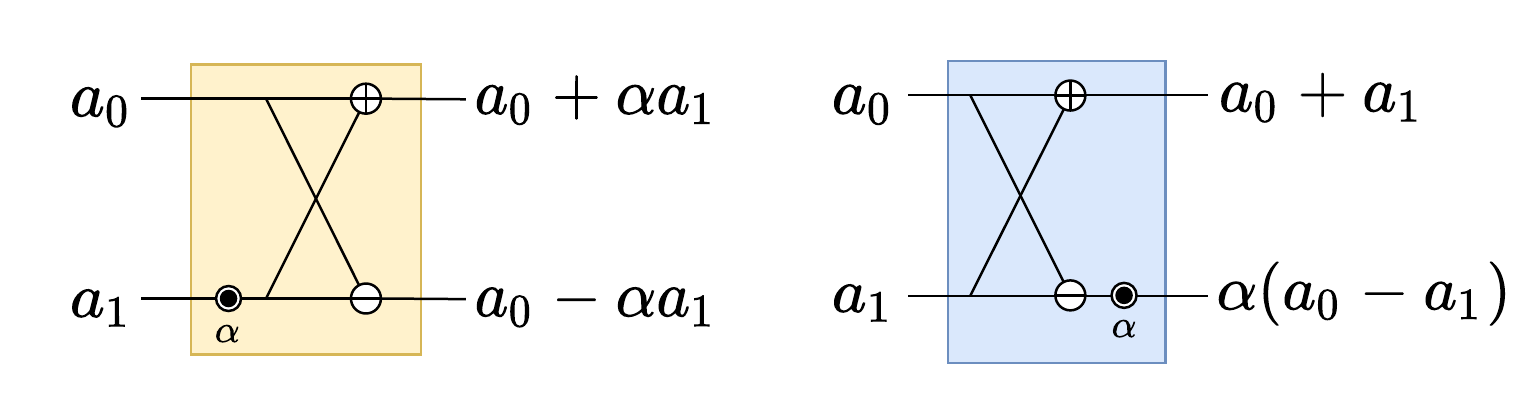}
	\caption{The Cooley--Tukey (left) and Gentleman--Sande butterflies (right).}
	\label{fig:ct_gs}
% \vspace{-1.75em}
\end{figure}

The NTT algorithm (i.e., the FFT) used to compute the NTT mathematical function (i.e., the DFT) consists of an iteration of \emph{stages}, in which computations are performed in the form of \emph{butterfly operations}.
The computational graphs for the well-studied (radix-2) Cooley--Tukey (CT) butterfly~\cite{cooleytukey} and the Gentleman--Sande (GS) butterfly~\cite{gentlemansande} are shown in Figure~\ref{fig:ct_gs}.  %\comment[id=Neal]{Note that the CT butterfly itself is a 2-point NTT (say precisely what this means) and that the GS butterfly is its inverse (say precisely what this means); maybe save this until fused negaconv section}

P{\"o}ppelmann et al.~\cite{poppelmann} define an elegant algorithmic specification for polynomial multiplication using NTTs based on the CT and GS butterflies. Their design utilizes two specialized variants of the FFT/NTT: 
\begin{enumerate}
\item the \emph{merged CT NTT}, $\mnttct{\psi}$, defined by Roy et al.~\cite{roy} (see Algorithm~\ref{alg:nttct2}); and 
\item the \emph{merged GS NTT}, $\mnttgs{\psi}$, defined by P{\"o}ppelmann et al.~\cite{poppelmann} (see Algorithm~\ref{alg:nttgs2}).
\end{enumerate}
\begin{algorithm}
\caption{Merged CT NTT, $\mnttct{\psi}$}\label{alg:nttct2}
\begin{algorithmic}[1]
\REQUIRE permuted twiddle factors $\roots_{\br}$
\STATE $m\gets 1$ \;
\STATE $k\gets \N/2$
\WHILE{$m<\N$}{
    \FOR{\texttt{$i=0$ to $m-1$}}{
        \STATE $jFirst\gets 2\times i\times k$ \;
        \STATE $jLast\gets jFirst + k - 1$ \;
        \STATE $\xi\gets\roots_{\br}[m+i]$ \;
        \FOR{\texttt{$j=jFirst$ to $jLast$}}{
            \STATE $\begin{bmatrix}\vect[j]\\ \vect[j+k]
            \end{bmatrix}\gets
    \begin{bmatrix}\vect[j]+\xi\times\vect[j+k]\Modq\\\vect[j]-\xi\times\vect[j+k]\Modq\end{bmatrix}$ \;
        }\ENDFOR
    }\ENDFOR
    \STATE $m\gets 2\times m$ \;
    \STATE $k\gets k/2$ \;
}\ENDWHILE
\RETURN $\vect$
\end{algorithmic}
\end{algorithm}
\begin{algorithm}
\caption{Merged GS NTT, $\mnttgs{\psi}$}\label{alg:nttgs2}
\begin{algorithmic}[1]
\REQUIRE permuted twiddle factors $\roots_{\br}$
\STATE $m\gets \N/2$ \;
\STATE $k\gets 1$
\WHILE{$m\ge 1$}{
    \FOR{\texttt{$i=0$ to $m-1$}}{
        \STATE $jFirst\gets 2\times i\times k$ \;
        \STATE $jLast\gets jFirst + k - 1$ \;
        \STATE $\xi\gets\roots_{\br}[m+i]$ \;
        \FOR{\texttt{$j=jFirst$ to $jLast$}}{
            \STATE $\begin{bmatrix}\vect[j]\\ \vect[j+k]
            \end{bmatrix}\gets
    \begin{bmatrix}\vect[j]+\vect[j+k]\Modq\\\xi\times(\vect[j]-\vect[j+k])\Modq\end{bmatrix}$ \;
        }\ENDFOR
    }\ENDFOR
    \STATE $m\gets m/2$ \;
    \STATE $k\gets 2\times k$ \;
}\ENDWHILE
\RETURN $\vect$
\end{algorithmic}
\end{algorithm}
In Algorithms~\ref{alg:nttct2} and \ref{alg:nttgs2}, $\br$ denotes the bit-reversal of a $\log_2(\N)$-bit binary sequence, and $\roots_{\br}$ denotes the twiddle factors permuted with respect to $\br$; i.e., $\roots_{\br}[i]=\psi^{\br(i)}$ for all $i$ in $[0,\N)$. %P{\"o}ppelmann et al.~\cite{poppelmann} show that p
Polynomial multiplication can be computed via the merged CT and GS NTTs as follows~\cite{poppelmann}:
\begin{equation}\label{eqn:poppelmann}
    \vect\negaconv\vecttwo = \frac{1}{\N}\mnttgs{\psi^{-1}}(\mnttct{\psi}(\vect)\odot\mnttct{\psi}(\vecttwo)).
\end{equation}
The advantages of this algorithmic specification for polynomial multiplication include the following:
\begin{enumerate}
\item \emph{Hadamard products omitted:} The multiplication by powers of $\psi$, i.e., the Hadamard products with  $\roots$ and $\rootsinv$, are ``merged'' into the NTT computations, saving a total of $3\N$ modular multiplications.
\item \emph{Bit-reversal permutations omitted:} The merged CT NTT takes the input in \emph{normal order} and returns the output in a permuted \emph{bit-reversed order} (hence $no\rightarrow bo$), and vice versa for the merged GS NTT. This removes the need for intermediate permutations to correct the order.
\item \emph{Good spatial locality:} In the merged CT NTT, the twiddle factors $\roots_{\br}$ are read in sequential order. In the merged GS NTT, the twiddle factors are read sequentially during each stage.
\end{enumerate}

Zhang et al.~\cite{zhang} propose a technique to merge the \mbox{ $\frac{1}{\N}$-scaling} operation in Equation (\ref{eqn:poppelmann}) into the GS NTT.
Rather than performing entry-wise modular multiplication by $\frac{1}{\N}$, Zhang et al. multiply the output of each butterfly operation by $\frac{1}{2}$ modulo $\q$. Observe that: 
\[\frac{x}{2}\Modq = 
\begin{cases} \frac{x}{2} &\text{if $x$ is even} \\ \floor{\frac{x}{2}} + \frac{q+1}{2} &\text{if $x$ is odd}\end{cases} \]
The computation of $\frac{x}{2}\Modq$ can be implemented without divisions, products, or branching via the expression \[x\gg 1 + (x\&1)\times ((q+1)\gg 1).\] 
\"{O}zerk et al.~\cite{ozerk} use this technique to merge $\frac{1}{\N}$-scaling into $\mnttgs{\psi^{-1}}$ (also see their open-source code \cite{ozerkcode}).
We write $\snttgs{\psi{-1}}$ to denote the merging of $\mnttgs{\psi^{-1}}$ with \mbox{$\frac{1}{\N}$-scaling}. Incorporating this NTT into (\ref{eqn:poppelmann}) gives the following algorithm specification for polynomial multiplication:
\begin{equation}\label{eqn:popzhang}
\vect\negaconv\vecttwo=\snttgs{\psi^{-1}}(\mnttct{\psi}(\vect)\odot\mnttct{\psi}(\vecttwo))\end{equation}
This algorithm specification is the basis for all of our implementations of polynomial multiplication.

% \subsection{Higher radix merged NTTs}
% Algorithmic specifications for a radix-4 CT NTT and GS NTT are given in Algorithm \ref{alg:nttct4} and Algorithm \ref{alg:nttgs4}, respectively. 

% \begin{algorithm}
% \caption{Merged radix-4 CT NTT}\label{alg:nttct4}
% \begin{algorithmic}[1]
% \REQUIRE $\N$ a power of $4$, $\vect\in(\Zq)^{\N}$

% permuted twiddle factors $\roots_{\perm{4}}\in(\Zq)^{\N}$
% \STATE $m\gets 1$ \;
% \STATE $k\gets \N/4$
% \WHILE{$m<\N$}{
%     \FOR{\texttt{$i=0$ to $m-1$}}{
%         \STATE $jFirst\gets 4\times i\times k$ \;
%         \STATE $jLast\gets jFirst + k - 1$ \;
%         \STATE $\ell\gets m+3\times i$
%         \STATE $\xi_0\gets \rootspermr[\ell]$ \;
%         \STATE $\xi_1\gets \rootspermr[\ell+1]$ \;
%         \STATE $\xi_2\gets \rootspermr[\ell+2]$ \;
%         \FOR{\texttt{$j=jFirst$ to $jLast$}}{
%             \STATE $\begin{bmatrix}\vect[j]\\ \vect[j+k]\\ \vect[j+2k]\\ \vect[j+3k]
%             \end{bmatrix}\gets
%     \begin{bmatrix}\vect[0]+\xi_0\times\vect[2]\Modq\\\vect[1]+\xi_0\times\vect[3]\Modq\\\vect[0]-\xi_0\times\vect[2]\Modq\\\vect[1]-\xi_0\times\vect[3]\Modq\end{bmatrix}$ \;
%             \STATE 
%             $\begin{bmatrix}\vect[j]\\ \vect[j+k]\\ \vect[j+2k]\\ \vect[j+3k]
%             \end{bmatrix}\gets
%     \begin{bmatrix}\vect[0]+\xi_1\times\vect[1]\Modq\\\vect[0]-\xi_1\times\vect[1]\Modq\\\vect[2]+\xi_2\times\vect[3]\Modq\\\vect[2]-\xi_2\times\vect[3]\Modq\end{bmatrix}$
%         }\ENDFOR
%     }\ENDFOR
%     \STATE $m\gets 4\times m$ \;
%     \STATE $k\gets k/4$ \;
% }\ENDWHILE
% \RETURN $\vect$
% \end{algorithmic}
% \end{algorithm}

\subsection{Proposed optimization: fused polynomial multiplication}
\label{sec:fused_ntt}
%\comment[id=Neal]{insert small introductory paragraph}
%\comment[id=Neal]{This next paragraph is meant to serve as an intro to the section. It needs polishing. I'm struggling to explain the ideas in this paragraph concisely. Maybe I should just omit some details? Or say something brief and write the rest in related work?}
Alkim et al.~\cite{alkim} propose several techniques for integrating the Hadamard product with its neighboring butterflies. They specify polynomial multiplication algorithms involving one, two, and three-stage integrations.
These algorithms have significantly reduced complexity for the multiplication of two polynomials. However, the complexity of multiplying larger numbers of polynomials may be significantly increased, especially when more stages are integrated. 

We propose a single-stage \emph{fused polynomial multiplication}, which offers significant speedup for multiplying two polynomials at minimized cost for multiplying larger numbers of polynomials. Our proposal uses Karatsuba's algorithm~\cite{karatsuba} to reduce the number of modular products by $\N/2$ compared to the single-stage algorithm of Alkim et al.~\cite{alkim}.

%Neal: TODO: translate the first minus sign to the right
% \begin{figure}[htbp]
% 	\centering
% 	\includegraphics[width=0.52\textwidth]{figures/04_polynomial_multiplication/fused_ntt_ab.pdf}
% 	\caption{(Top) The Hadamard product and its neighboring butterflies (Bottom) Fusion of butterflies into Hadamard product}
% 	\label{fig:fused_ntt}
% \end{figure}

% \begin{figure}[htbp]
% 	\centering
% 	\includegraphics[width=0.45\textwidth]{figures/04_polynomial_multiplication/fused_ntt_b.pdf}
% 	\caption{Fusion of butterflies into Hadamard product}
% 	\label{fig:fused_ntt}
% \end{figure}

Consider the computational subgraph of Equation (\ref{eqn:popzhang}) induced by the final stages of $\mnttct{\psi}$, the Hadamard product $\odot$, and the first stage of $\snttgs{\psi^{-1}}$. Each of the $\N/2$ connected components in this graph are of the form
% $\buttgs{\rootsinv[\N/2+i]}\left(\buttct{\roots[\N/2+i]}(\widehat{\vect}[2i], \widehat{\vect}[2i+1])
% %\begin{bmatrix} \vect_{2i}\\\vect_{2i+1}\end{bmatrix}
% \odot\buttct{\roots[\N/2+i]}
% %\begin{bmatrix} \vecttwo_{2i}\\\vecttwo_{2i+1}\end{bmatrix}
% (\widehat{\vecttwo}[2i], \widehat{\vecttwo}[2i+1])
% \right)$
\begin{equation}\label{eqn:butt} \frac{1}{2}\buttgs{\alpha^{-1}}\left(\buttct{\alpha}\left(
\begin{bmatrix}a_0\\a_1\end{bmatrix}\right)
\odot\buttgs{\alpha}
\left(\begin{bmatrix} b_0\\b_1\end{bmatrix}\right)\right)\end{equation}
for some twiddle factor $\alpha$ and inputs $a_0,a_1,b_0$, and $b_1$ (see Figure~\ref{fig:fused_ntt}).
Thus, the computation for each component consists of 5 (modular) product operations, 2 scaling by $\frac{1}{2}$ operations, 6 sum/difference operations, and $2$ memory accesses.
The output of the computation in expression (\ref{eqn:butt}) is
\begin{equation}\label{eqn:fusedbutt} \begin{bmatrix} a_0\times b_0 + \alpha^2\times a_1\times b_1\Modq  \\a_0\times b_1 + a_1\times b_0\Modq\end{bmatrix}\end{equation}

Algorithm~\ref{alg:fusedbutt} also computes expression (\ref{eqn:butt}), but requires 4 products, 0 scalings by $\frac{1}{2}$, 5 sums/differences, and $1$ memory access. This variation on Karatsuba's Algorithm~\cite{karatsuba} relies on the fact that $a_0\times b_1 + a_1\times b_0 \Modq$ is equivalent to 
\[ (a_0+a_1)\times(b_0+b_1) - a_0\times b_0 - a_1\times b_1\Modq.\] We say that Algorithm~\ref{alg:fusedbutt} \emph{fuses} the CT and GS butterflies into the Hadamard product.

\begin{algorithm}
\caption{Butterflies fused into the Hadamard product}\label{alg:fusedbutt}
\begin{algorithmic}[1]
\REQUIRE $\left[\begin{smallmatrix} a_0 \\ a_1 \end{smallmatrix}\right],\left[\begin{smallmatrix} b_0 \\ b_1 \end{smallmatrix}\right]\in(\Zq)^2$, twiddle factor $\alpha^2\in\Zq$
\ENSURE $\left[\begin{smallmatrix} c_0 \\ c_1 \end{smallmatrix}\right]\gets \left[\begin{smallmatrix} a_0\times b_0 + \alpha^2\times a_1\times b_1\Modq  \\a_0\times b_1 + a_1\times b_0\Modq\end{smallmatrix}\right]$
\STATE $prod1\gets a_0\times b_0\Modq$ \;
\STATE $prod2\gets a_1\times b_1\Modq$ \;
\STATE $sum1\gets a_0+a_1\Modq$ \;
\STATE $sum2\gets b_0+b_1\Modq$ \;
\STATE $prod3\gets sum1\times sum2\Modq$ \;
\STATE $prod4\gets \alpha^2\times prod2\Modq$ \;
\STATE $sum3\gets prod1 + prod4\Modq$ \;
\STATE $sum4\gets prod3 - prod1\Modq$ \;
\STATE $sum5\gets sum4 - prod2\Modq$ \;
\STATE $\left[\begin{smallmatrix} c_0 \\ c_1 \end{smallmatrix}\right]\gets\left[\begin{smallmatrix} sum3 \\ sum5 \end{smallmatrix}\right]$
\RETURN $\left[\begin{smallmatrix} c_0 \\ c_1 \end{smallmatrix}\right]$
\end{algorithmic}
\end{algorithm}

% Note that the twiddle factors read in the final stage of $\mnttct{\psi}$ are $\psi^{\br(\N/2+i)}\Modq$ for $i$ in $[0,\N/2)$. If $i$ lies in $[0,\N/2)$, then
% \[(\psi^{\br(\frac{\N}{2}+i)})^2\Modq=
% \begin{cases}
% \psi^{\br(\frac{\N}{4}+\frac{i}{2})}\Modq & \text{if $i$ is even}\\
% -\psi^{\br(\frac{\N}{4}+\frac{i-1}{2})}\Modq & \text{if $i$ is odd}
% \end{cases}\]
% Thus the squares of the twiddle factors in the last stage (i.e., $\psi^{\br(i)}\Modq$ for $i$ in $[\frac{\N}{2},\N)$) are the twiddle factors in the second to last stage (i.e., $\psi^{\br(i)}\Modq$ for $i$ in $[\frac{\N}{4},\frac{\N}{2})$) up to a $+$ or $-$ sign.
%Thus in line 2 of Algorithm~\ref{alg:fusedhadamard}, $\gamma=(\psi^{\br(\N/2+i)})^2\Modq$. 
% double-check your idea about squares of second half of twiddles being second quarter with negative signs

To define the fused polynomial multiplication algorithm, we first define truncated versions of the CT and GS NTTs.
Define the \emph{truncated CT NTT}, $\tnttct{\psi}$, to be the merged CT NTT with the final stage omitted (i.e., line 3 in Algorithm~\ref{alg:nttct2} is replaced with ``$\text{{\bf while} $m<(\N/2)$ {\bf do}}$'').
Likewise, define the \emph{truncated GS NTT}, $\tnttgs{\psi}$, to be the merged GS NTT with the first stage omitted (i.e., line 3 in Algorithm~\ref{alg:nttgs2} is replaced with ``$\text{{\bf while} $m>1$ {\bf do}}$''). Our proposed \emph{fused polynomial multiplication} is specified in Algorithm~\ref{alg:fusedhadamard}.
%A specification for the full \emph{fused Hadamard product}, i.e., the disjoint union of the fusions of each of the $\N/2$ connected components, is defined in Algorithm \ref{alg:fusedhadamard}. We should note that lines 8--11 can be implemented without branching. For example, if $\q<\beta/4$, lines 8--11 can be replaced with \[\vectthree[2i]\gets x+z+(i\&1)\times((\q-z)\ll 1)\Modq.\]
\begin{algorithm}
\caption{Proposed fused polynomial multiplication}\label{alg:fusedhadamard}
\begin{algorithmic}[1]
\REQUIRE $\vect,\vecttwo\in(\Zq)^{\N}$, permuted twiddle factors $\roots_{\br}$
\ENSURE $\vectthree=\vect\negaconv\vecttwo$
\STATE $\vectimg = \tnttct{\psi}(\vect)$ \;
\STATE $\vecttwoimg = \tnttct{\psi}(\vecttwo)$ \;
\FOR{\texttt{$i=0$ to $\N/2-1$}}{
%    \STATE $\gamma\gets (-1)^i\times\roots_{\br}[\frac{\N}{4}+\floor{\frac{i}{2}}]\Modq$ \;
    \STATE $u \gets \vectimg[2i]\times\vecttwoimg[2i]\Modq$ \;
    \STATE $v \gets \vectimg[2i+1]\times\vecttwoimg[2i+1]\Modq$ \;
    \STATE $w\gets(\vectimg[2i]+\vectimg[2i+1])\times(\vecttwoimg[2i]+\vecttwoimg[2i+1])\Modq$ \;
    \STATE $y\gets w - u\Modq$ \;
    \STATE $\vectthreeimg[2i+1]\gets y - v\Modq$ \;
    \STATE $z\gets v\times \roots_{\br}[\frac{\N}{4}+\floor{\frac{i}{2}}] \Modq$ \;
    \IF{\texttt{$i$ is even}}{
        \STATE $\vectthreeimg[2i]\gets u + z\Modq$ \;
    }\ELSE{
        \STATE $\vectthreeimg[2i]\gets u - z\Modq$ \;
    }\ENDIF
}\ENDFOR
\STATE $\vectthree \gets \tnttgs{\psi^{-1}}(\vectthreeimg)$ \;
\RETURN $\vectthree$
\end{algorithmic}
\end{algorithm}

%  We propose the following \emph{fused polynomial multiplication} algorithm:
% \begin{equation}\label{eqn:fusedpoly}
%     \vect\negaconv\vecttwo = \tnttgs{\psi^{-1}}(\tnttct{\psi}(\vect)\fusedhada\tnttct{\psi}(\vecttwo)).
% \end{equation}
The benefits of our proposed fused polynomial multiplication algorithm include the following:
\begin{enumerate}
\item \emph{Fewer operations:} $\N/2$ fewer modular product operations, $\N$ fewer scaling by $\frac{1}{2}$ operations, $\N/2$ fewer sum/difference operations, and $\N/2$ fewer memory accesses (but $\N/4$ additional negations).
\item \emph{Number of twiddle factors halved:} The second half of the entries in the twiddle factor arrays for each of the merged NTTs are not used in fused polynomial multiplication and can be omitted.
\item \emph{Re-use of recently-accessed twiddle factors}: The twiddle factors read in the last stage of the truncated CT NTT are immediately re-used in the fused Hadamard product.
\end{enumerate}
%However $\N/4$ negation operations are required in our proposed fused polynomial multiplication algorithm. These are operations that are not required in Equation (\ref{eqn:popzhang}). 
%The performance analysis of fused polynomial multiplication is given in Section \comment[id=Neal]{refer to section on performance analysis}

%Although Alkim et al. improve the performance of (a single) polynomial multiplication by further truncating their NTT's, the output of these more heavily truncated NTT's is further removed from the ``frequency domain'' (i.e., the usual codomain of the NTT).  requires more operations to transform the output of a heavily truncated NTT to the usual NTT output. Although some of Alkim et al.'s algorithms utilize Karatsuba's Algorithm, they do not consider using Karatsuba's Algorithm to merge a single innermost pair of NTT stages. In our tests, our fused polynomial multiplication implementation out-performs our implementation of polynomial multiplication using Alkim et al.'s ``$(k-1)$-level NTT then Schoolbook Multiplication'' algorithm by \comment[id=Neal]{$XX\%$}.

%   ____ ____  _   _ 
%  / ___|  _ \| | | |
% | |  _| |_) | | | |
% | |_| |  __/| |_| |
%  \____|_|    \___/ 

\section{GPU Architecture}
\label{sec:gpu}
% \section{Evaluation Methodology}
% Table of papers
% Extrapolate results for resnet
% Examples of post quantum applications
%The GPU is a Single Instruction Multiple Thread (SIMT) style architecture that allows for running parallel workloads. 
We implement our polynomial multiplication kernels targeting NVIDIA's $7^{th}$ generation Volta GPU architecture, the V100 PCIe GPU with 16 GB onboard memory.
% This section highlights the elements of this GPU that we consider in our NTT implementation. 
%\subsection{Compute Elements}
%Workloads that run on GPU have to be programmed as a single program multiple data (SPMD) style kernel.
%The V100 GPU consists of $84$ streaming multiprocessors (SMs).
%Each SM consists of $64$ INT32 cores capable of performing integer arithmetic operations, totaling to $5376$ INT32 cores per GPU.
%The GPU executes kernels in a grid of blocks, where each block is run on a single SM.  Within a SM, the kernel is run in a lockstep manner, in groups of 32 threads called a warp.
%\subsection{V100 Memory Hierarchy}
\begin{figure}[b]
	\centering
	\includegraphics[width=0.48\textwidth]{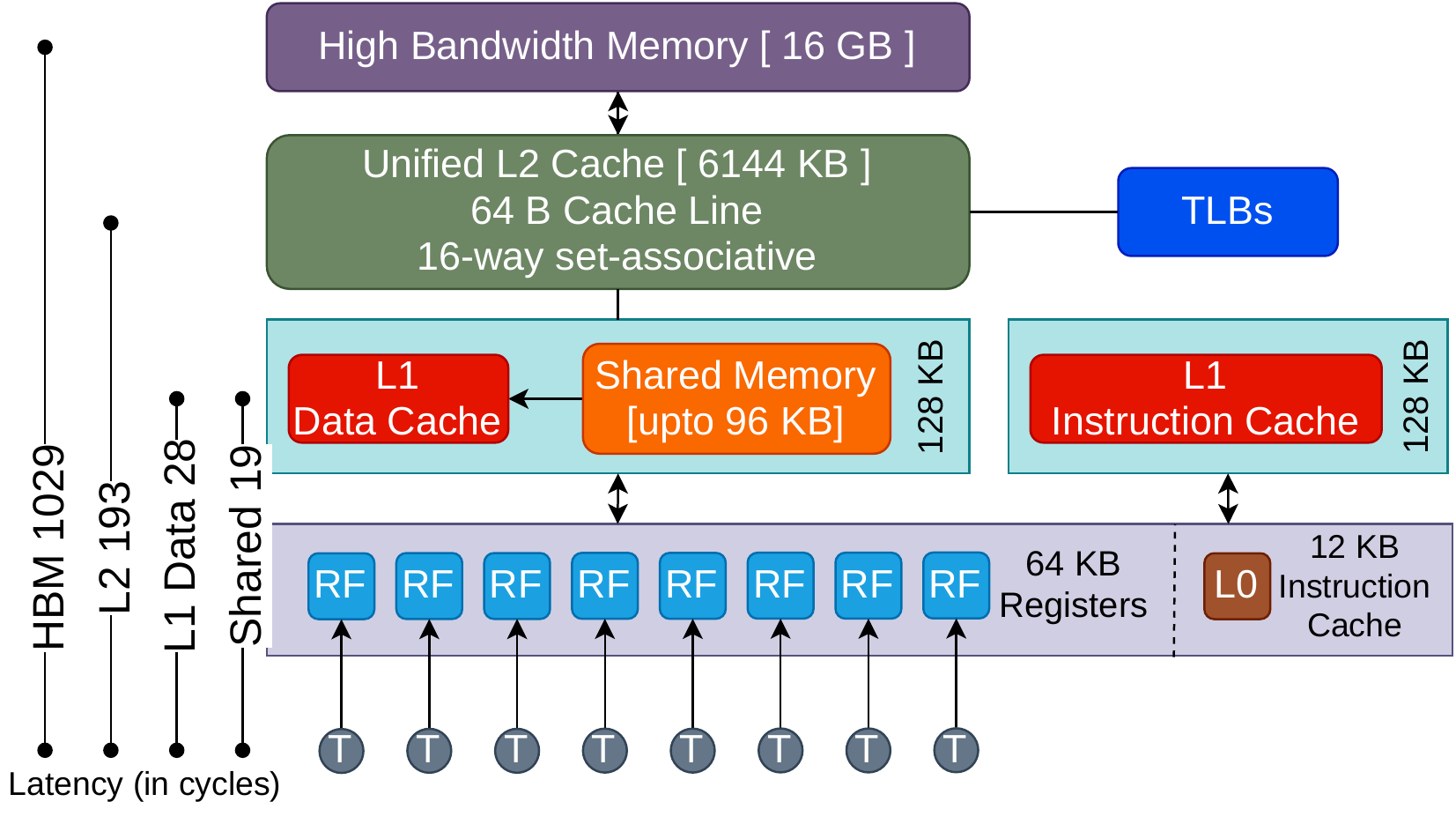}
	\caption{V100 GPU memory hierarchy and latency comparison.} 
%(Assuming conflict-free accesses for latency values.)}
	\label{fig:gpu_memory}
\vspace{-0.5em}
\end{figure}
The V100 has a multi-level memory, as shown in Figure~\ref{fig:gpu_memory}.  The V100 features a highly tuned high-bandwidth memory (HBM2), which is called global memory in the CUDA framework. The global memory, being the largest in capacity, has the highest latency to access data ($\mathtt{\sim}1029$ cycles)~\cite{jia2018dissecting}. 
%The V100 has a register file as the fastest storage tier. Each register file is divided into $2$ banks of 64-bit registers~\cite{jia2018dissecting}.
%These wide register banks allow the compiler to generate conflict-free code~\cite{jia2018dissecting}. 
% The fastest data storage available to the threads are registers.
%Each thread is allotted 255 registers, totaling 65536 registers in the register file per streaming multiprocessor (SM). Data can be read from registers by the threads in a single cycle (when memory transactions are free from bank conflicts). 
The V100 provides a $128$ KB L1 data cache and a $128$ KB L1 instruction cache per SM, as well as a unified L2 cache for data and instructions ($6.1$ MB in size). Each SM on a V100 has a shared memory (each configurable in size up to 96 KB). Data accesses to shared memory are much more efficient (i.e., $\mathtt{\sim}19$ cycles) as compared to accesses to global memory ($1029$ cycles)~\cite{baruah2021gnnmark}.
% Shared memory can only be accessed by threads within a thread block.
Effective use of the memory hierarchy, and especially shared memory, on a GPU is critical to obtaining the best performance~\cite{sun2019mgpusim}.  Our single-block implementation of NTT utilizes shared memory for local data caching, thus reducing the number of redundant fetches from global memory~\cite{hashing2021} by a factor of $\log(\N)$ times (where $\N$ is the size of the input coefficient array). 
%(Further described in \S\ref{section:shared_memory}).
%Each V100 global memory transaction provides 128 bytes of data, regardless of the number of bytes required. Ideally, if 32 threads of a warp access 4 bytes of cache-aligned contiguous elements, then all the data fetched by a single memory transaction would be utilized (thus improving L2 and L1 cache efficiency).
Furthermore, we improve cache efficiency by increasing the spatial locality of our data access patterns, exploiting memory coalescing on the GPU~\cite{hashing2021}, as described in Section~\ref{section:tom_ntt}.

% \textcolor{blue}{We present $3$ NTT kernels in this paper along with $4$ iterative optimizations tailored for the GPU platform.
% We evaluate the performance of our single-block NTT kernel for input coefficient vector size of $N=2^{11}$ and the performance of our multi-block kernel for vector size $N=2^{12}~\mathrm{to}~2^{16}$.
% We iteratively add the $4$ optimizations to our NTT kernels and report performance improvements. Twiddle factors are pre-computed on the CPU and hence do not add compute overhead on the GPU. We report the GPU kernel execution time for each approach.
% For each optimization, the speedup achieved is reported using the respective non-optimized kernel as the baseline for comparison.
% Finally we evaluate the scale up properties by performing weak scaling experiments on our throughput optimized TOM-NTT kernel.}

We obtain performance metrics for our kernels using hardware performance counters and binary instrumentation tools. We explore performance bottlenecks using a variety of tools including the NVIDIA Binary Instrumentation Tool (NVBit)~\cite{villa2019nvbit} for tracing memory transactions, the Nsight Compute for fetching performance counters, and the Nsight Systems~\cite{nsight_systems} to obtain kernel scheduler performance, as well as measuring synchronization overheads.
We compare kernel performance based on ``Architectural Profile'' and ``Stall Profile'' plots. The ``Arch Profile'' compares the relative change as compared to a baseline (see Table~\ref{table:arch_profile}), whereas the ``Stall Profile'' provides information on the primary causes of a kernel stall during execution (see Table~\ref{table:stall_profile}).

%% Can be pushed to APPENDIX if we run out of space in paper.
\begin{table}[htbp]
% \vspace{-0.6em}
	\centering % used for centering table
	% p{20mm}
	\begin{tabular}{l l} % centered columns (4 columns)
		\hline\hline %inserts double horizontal lines
		\textbf{Parameter} & \textbf{Description} \\ [1ex]
		%heading
		\hline %\\ [0.5ex] % inserts single horizontal line
		SM Throughput & \% of cycles the SM was busy \\ [1ex]
		\hline
		Avg. IPC & Average \# of instructions per cycle \\ [1ex]
		\hline
		ALU & ALU Pipeline utilization \\ [1ex]
		\hline
		DRAM B/W & \begin{tabular}{@{}l@{}} \% of peak memory transactions \\ the DRAM processed per second \end{tabular} \\ [1ex]
		\hline
		L1\$ and L2\$ B/W & \begin{tabular}{@{}l@{}} \% of peak memory transactions the L1\$ \\ and L2\$ processed per second respectively \end{tabular} \\ [1ex]
		\hline
		L1\$ and L2\$ Hit-Rate & \begin{tabular}{@{}l@{}} \% of memory transactions the L1\$ \\ and L2\$ fulfilled successfully \end{tabular} \\ [1ex]
		\hline
		Regs/Thread & \# of registers used by each thread of the warp \\ [1ex]
		\hline
		Issued Warps & Avg. \# of warps issued per second by scheduler\\ [1ex]
		%\hline % inserts single horizontal line
		%Fused Hadamard Product$^*$ & $1.31\times$ & $+9.9\%$ & $+18.0\%$ \\ [1ex]
		\hline \\ [1ex]  %inserts single line 
	\end{tabular}
	\caption{Description of the Arch Profile parameters.} % title of Table
	\label{table:arch_profile} % is used to refer this table in the text
\vspace{-1.8em}
\end{table}

%% Can be pushed to APPENDIX if we run out of space in paper.
\begin{table}[htbp]
% \vspace{-0.6em}
	\centering % used for centering table
	% p{20mm}
	\begin{tabular}{l l} % centered columns (4 columns)
		\hline\hline %inserts double horizontal lines
		\textbf{Type of stall} & \textbf{Reason} \\ [1ex]
		%heading
		\hline %\\ [0.5ex] % inserts single horizontal line
		Long Scoreboard & \begin{tabular}{@{}l@{}} Waiting for a scoreboard dependency \\ on a L1\$ operation \end{tabular} \\ [1ex]
		\hline
		Math Pipe Throttle & \begin{tabular}{@{}l@{}} Waiting for the ALU execution pipe \\ to be available \end{tabular} \\ [1ex]
		\hline
		Wait & \begin{tabular}{@{}l@{}} Waiting on fixed latency execution dependency \\ Indicates highly optimized kernel \end{tabular} \\ [1ex]
		\hline
		Not Selected & \begin{tabular}{@{}l@{}} Waiting for the scheduler to select the warp \\ Indicates warps oversubscribed to scheduler \end{tabular} \\ [1ex]
		\hline
		Selected & Warp was selected by the micro scheduler\\ [1ex]
		\hline
		Barrier & \begin{tabular}{@{}l@{}} Waiting for sibling warps at sync barrier \\ Indicates diverging code paths before a barrier \end{tabular} \\ [1ex]
		\hline
		LG Throttle & Waiting for the L1 instruction queue \\ [1ex]
		\hline
% 		\begin{tabular}{@{}l@{}} A \\ B \end{tabular} \\ [1ex]
		Short Scoreboard & \begin{tabular}{@{}l@{}} Scoreboard dependency on shared memory \\ Indicates higher shared memory utilization \end{tabular} \\ [1ex]
		\hline
		MIO Throttle & Stalled on MIO (memory I/O) instruction queue \\ [1ex]
		\hline
		Branch Resolving & Waiting for a branch target to be computed \\ [1ex]
		\hline
		Dispatch Stall & \begin{tabular}{@{}l@{}} Warp stalled because dispatcher holds back \\ issuing due to conflicts or events \end{tabular} \\ [1ex]
		\hline
		IMC Miss & Waiting for an immediate cache (IMC) miss \\ [1ex]
		\hline
		No Instruction & Waiting after an instruction cache miss \\ [1ex]
		%\hline % inserts single horizontal line
		%Fused Hadamard Product$^*$ & $1.31\times$ & $+9.9\%$ & $+18.0\%$ \\ [1ex]
		\hline \\ [1ex]  %inserts single line 
	\end{tabular}
	\caption{Description of the Stall Profile parameters.} % title of Table
	\label{table:stall_profile} % is used to refer this table in the text
\vspace{-2.5em}
\end{table}

%                 _       _   _ _____ _____ 
%   ___ _   _  __| | __ _| \ | |_   _|_   _|
%  / __| | | |/ _` |/ _` |  \| | | |   | |  
% | (__| |_| | (_| | (_| | |\  | | |   | |  
%  \___|\__,_|\__,_|\__,_|_| \_| |_|   |_|  

\section{Optimized NTT Kernels}

\begin{figure*}[htbp]
	\centering
	\includegraphics[width=1.0\textwidth]{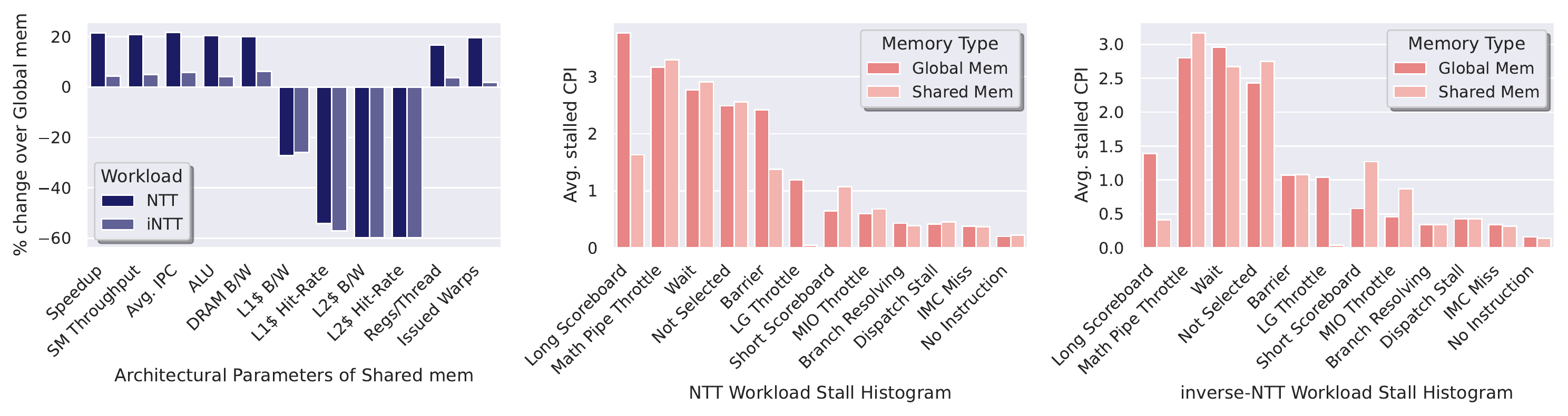}
	\vspace{-2em}
	\caption{(a) Architectural performance profile of shared memory NTT and iNTT workloads compared against respective global memory workloads. (b) Stall profile of NTT workload comparing global and shared memory kernels. (c) Stall profile of inverse-NTT workload comparing global vs. shared memory kernels.}
	\label{fig:memory_profile}
\vspace{-1.5em}
\end{figure*}

We observe that the NTT kernel is a memory-bound workload, heavily bottlenecked by the GPU's DRAM latency. The butterfly operation is one of the key computations within the NTT kernel. This operation is characterized by strided accesses, with the stride varying with each stage.
%(as seen in Figure~\ref{fig:los_ntt_wire} \comment[id=KTB]{Add LOS-NTT wire diagram}).
The changes in the stride lead to non-sequential memory accesses, reducing the spatial locality of the NTT kernel. To effectively leverage memory coalescing, we can partition data carefully across CUDA threads~\cite{speeding_gpu}.
%This in-turn would lead to an improvement on the spatial locality present in the data access patterns within the kernel.
%In this paper we present various polynomial multiplication kernels referred to as PEACHES (\textbf{P}olynomial-Multiplication \textbf{A}cceleration on \textbf{C}UDA enabled GPUs for \textbf{H}omomorphic \textbf{E}ncryption \textbf{S}chemes).
We propose three different implementations of NTT kernels, each optimized for different input sizes and employing different data partitioning techniques.  We follow a similar approach here as described by {\"O}zerk et al.~\cite{ozerk}, though we leverage a number of algorithmic optimizations, combined with code optimizations, that are unique to this work.

The three implementations of polynomial multiplications proposed in this work are as listed below:
\begin{itemize}
    \item \emph{LOS-NTT} (Latency-optimized Single-block NTT): For single polynomial multiplication with $\N~\leq~2^{11}$.
    \item \emph{LOM-NTT} (Latency-optimized Multi-block NTT): For single polynomial multiplication with $\N~>~2^{11}$.
    \item \emph{TOM-NTT} (Throughput-optimized Multi-block NTT): For multiple polynomial multiplications with no constraints on $\N$.
\end{itemize}

\subsection{Latency optimized Single-block NTT}

%  a cache hit reduces DRAM bandwidth demand but not fetch latency
The LOS-NTT kernel performs all the NTT operations within a single block of the CUDA kernel.  Using a single block for computing the entire NTT workload has the following advantages:
\begin{itemize}
    \item The overhead of a single block-level barrier ($syncthreads$) is significantly lower than a kernel-level (multi-block) barrier.
    \item We can leverage shared memory, which can only be addressed within the scope of a single block. 
    \item Since all threads of a block share the same L1 and L2 caches, and L1 is write-through, write updates by any thread are reflected in L2 across all threads.
\end{itemize}

% Our LOS-NTT kernel is implemented using the merged NTTs defined in Algorithms~\ref{alg:nttct2} and \ref{alg:nttgs2}.
Our LOS-NTT implementation consists of two phases, separated by a block-level barrier. The first transfers the input coefficient vectors (of size $\N$) from high latency global memory to the faster, low latency shared memory.
The second phase performs the merged NTTs, as defined in Algorithms~\ref{alg:nttct2} and~\ref{alg:nttgs2}.
This phase consists of two nested loops. The first iterates over the $\log(\N)$ stages of the CT algorithm. This is followed by the second loop of $\frac{\N}{2}$, iterating over the elements of the input coefficient vector.  These iterations are free from any loop-carried dependencies, allowing them to be run in parallel. We capitalize on this inherent parallelism by computing each iteration of the second loop in parallel, assigning each loop iteration to a separate CUDA thread. We further improve the performance of our kernel with four GPU-specific optimizations.

\subsubsection{Shared Memory Optimization}
\label{section:shared_memory}

Each stage of the CT implementation is characterized by multiple butterfly operations of varying strides.  These butterfly operations result in strided memory accesses, with the step size varying from $1$ to $\frac{\N}{2}$ (where $\N$ can be as large as $2^{16}$). We optimize for memory access efficiency by storing the input coefficient vector, as well as the outputs of butterfly operations, in persistent shared memory, which is significantly faster than accessing global memory.  We utilize $8$ KB of shared memory per SM for storing the input polynomial coefficients, as well as the output of intermediate stages.
Using shared memory incurs the overhead of transferring input coefficients to shared memory and the final results back to global memory.
Despite these additional overheads, incorporating the use of shared memory allows us to obtain a $1.25\times$ speedup over the use of only global memory (Figure~\ref{fig:memory_profile}).  Figure~\ref{fig:memory_profile}(a) denotes a large drop in $L1\$$ and $L2\$$ performance. The primary reason for this degraded performance is memory transactions that access the shared memory do not count towards $L1$ and $L2$ cache performance. Since all the coalesced memory transactions to the global memory (which counted towards the cache performance) are now redirected towards the shared memory (which is excluded from cache performance counters), the $L1$ and $L2$ cache bandwidth and hit-rate take a performance hit. Figure~\ref{fig:memory_profile}(b,c) identifies the primary causes of stalls for the NTT and inverse-NTT kernels, respectively.  The``Long Scoreboard'' stall is caused by dependencies in L1 cache operations. The large drop in the stall values for the ``Long Scoreboard'' in Figure~\ref{fig:memory_profile}(b) is an indicator of memory pressure being reduced in the L1 cache and indirectly in the L2 cache and DRAM.
Similarly, in Figure~\ref{fig:memory_profile}(c), the increase in the average ``Math Pipe Throttle'' stall values is tied to the compute throughput of the inverse-NTT kernel.

% \begin{figure}[ht]
% % \vspace{-0.5em}
% 	\centering
% 	\includegraphics[width=0.48\textwidth]{figures/08_results/memory/shared_single_sns.pdf}
% 	\caption{Arch. Profile to visualize persistent shared memory optimization normalized to the performance of using only global memory.}
% %\vspace{-0.9em}
% 	\label{fig:single_block_ntt_timing}
% % \vspace{-0.8em}
% \end{figure}

% \begin{figure}[ht]
% \vspace{-0.5em}
% 	\centering
% 	\includegraphics[width=0.48\textwidth]{figures/08_results/memory/mem_stall_ntt_sns.pdf}
% 	\caption{Stall profile for NTT}
% %\vspace{-0.9em}
% 	\label{fig:single_block_ntt_timing}
% \vspace{-0.8em}
% \end{figure}

% \begin{figure}[ht]
% \vspace{-0.5em}
% 	\centering
% 	\includegraphics[width=0.48\textwidth]{figures/08_results/memory/mem_stall_intt_sns.pdf}
% 	\caption{Stall profile inverse-NTT}
% %\vspace{-0.9em}
% 	\label{fig:single_block_ntt_timing}
% \vspace{-0.8em}
% \end{figure}

\subsubsection{Barrett's Modular Reduction Optimization}
We further accelerate our NTT kernel with the use of our modified Barrett implementation, specifically designed for GPU execution, as shown in Section~\ref{sec:barrett}. The smaller number of correctional subtractions present in our implementation allows us to obtain a $1.85\times$ average speedup over previous work~\cite{ozerk} and a $1.72\times$ speedup over the builtin modulus operation.
We also obtain similar execution times to the $28$-bit modified Barrett's reduction, as reported for PALISADE~\cite{palisade}.
To our knowledge, our proposed Barrett variant is the fastest Barrett modular reduction for general $30$-bit and $62$-bit moduli.
% Our optimized implementation of Barrett's modular reduction, when incorporated in polynomial multiplication, achieves a $2.8\%$ improvement over polynomial multiplication that utilizes classical Barrett's reduction.

% \subsubsection{Higher Radix Optimization}
% \begin{figure}[htbp]
% 	\centering
% 	\includegraphics[width=0.485\textwidth]{figures/04_polynomial_multiplication/radix_4.png}
% 	\caption{Radix-4 butterfly implementation.}
% 	\label{fig:radix_4_wire}
% \end{figure}
\subsubsection{Mixed Radix Optimization}
The naive implementation of NTT and inverse NTT used in this study is based on a radix-$2$ algorithm.
%as seen in Figure~\ref{fig:los_ntt_wire}.
In this implementation, each thread within a block operates on $2$ elements of the input coefficient array.  We improve the performance of our kernel by experimenting with radix-$4$, $8$, and $16$ implementations that distribute $4$, $8$, and $16$ elements per thread, respectively.  Higher radix implementations improve temporal locality, as the input coefficient vector data is reused.  Unfortunately, this improvement in the temporal locality is associated with a significant loss in parallelism.
%This can be observed in Table~\ref{table:radix_parallelism}.
%As an example, the lower parallelism in radix-$16$ limits the number of parallel tasks that can be launched simultaneously by a factor of $16$.
We also experiment with kernels that use radix $4$ or radix $8$ for single-block kernels and radix $16$ for multi-block.

\begin{figure}[htb]
\vspace{-0.8em}
	\centering
	\includegraphics[width=0.48\textwidth]{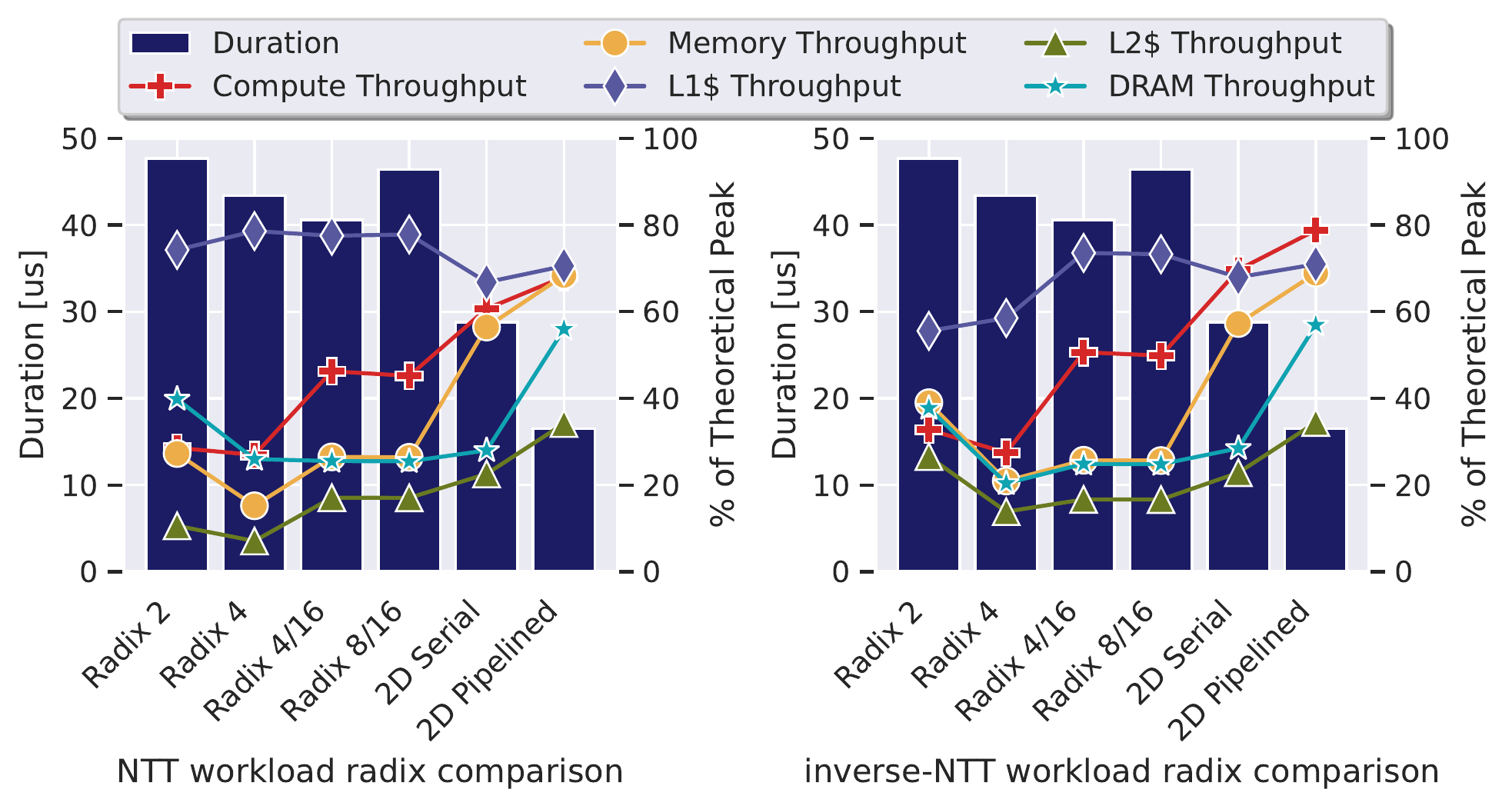}
	\vspace{-1.0em}
	\caption{Higher radix comparison for (a) NTT and (b) inverse-NTT kernels.}
	\label{fig:radix_timing}
\vspace{-0.8em}
\end{figure}

We also experiment with $2$-dimensional NTT implementations.
% to help us distribute the workload to increase the parallelism.
A 2D NTT maps the data into a matrix form, thus treating our coefficient as a row-major square matrix. This allows us to perform a column-wise NTT followed by a row-wise NTT. An $N-1$ degree polynomial can be mapped into a $\sqrt{N} \times \sqrt{N}$ matrix. This also divides the NTT kernel into two stages (column-wise NTT and row-wise NTT). The first stage computes $\sqrt{N}$ number of $\sqrt{N}$-point column-wise NTT operations, followed by the second stage that computes $\sqrt{N}$ number of $\sqrt{N}$-point row-wise NTT. Each $\sqrt{N}$-point NTT is mapped into a block with $\frac{\sqrt{N}}{2}$ threads, where each thread is responsible for computing a radix-$2$ butterfly.
% By dividing the work this way we only need to synchronize between blocks once.
The 2D NTT approach allows us to map the data while preserving spatial locality. We further accelerate our computation by pipelining the two stages of row-wise and column-wise NTT operations, thus presenting two variants of our $2D$ implementation ($2D$ Serial and $2D$ Pipelined). This approach provides an average of $2.91\times$ speedup for NTT and inverse-NTT kernel over the naive radix-$2$ implementations (Figure~\ref{fig:radix_timing}). This improvement in execution time can be largely attributed to the increased memory throughput for NTT (Figure~\ref{fig:radix_timing}(a)), as well as inverse-NTT (Figure~\ref{fig:radix_timing}(b)). The improved memory throughput also contributes to the increased compute throughput, as continuous streaming of data from DRAM no longer starves the SMs of input operands.
% In the previous approach, as we increase the radix, the number of butterfly operations inside 1 thread will increase. However, at some point we can map 1 high radix butterfly operation into 1 block instead of 1 thread, which leads to this 2D NTT implementation.

\subsubsection{Fused Polynomial Multiplication Optimization}
Finally, we propose an optimization that fuses together the last stage of merged CT NTT, the Hadamard product, and the first stage of merged GS NTT.  Figure~\ref{fig:fused_ntt}(c) shows the implementation of our fused polynomial multiplication.  Our implementation of the fused kernel significantly reduces the number of multiplication operations and re-uses recently-cached twiddle factors.
% The naive implementation utilizes $XX$ multiplications and $XX$ additions. Our fused Hadamard product kernel incorporates $XX$ multiplications and $XX$ additions.
Experimental results show that we reduce the execution time,
%from $8.51~\mu\mathrm{s}$ down to $5.82~\mu\mathrm{s}$,
resulting in a $6.1\%$ and $2.4\%$ improvement as compared to the naive implementation for polynomial multiplication, for input sizes of $N=2^{11}$ and $N=2^{16}$, respectively.

\subsection{Latency optimized Multi-block NTT}
% \begin{figure}[htbp]
% \vspace{-1.2em}
% 	\centering
% 	\includegraphics[width=0.48\textwidth]{figures/06_cudantt/lom_ntt.pdf}
% 	\caption{Multi-Block NTT task distribution}
% 	\label{fig:multi_block}
% \end{figure}

\begin{figure}[htbp]
\vspace{-1.2em}
	\centering
	\includegraphics[width=0.48\textwidth]{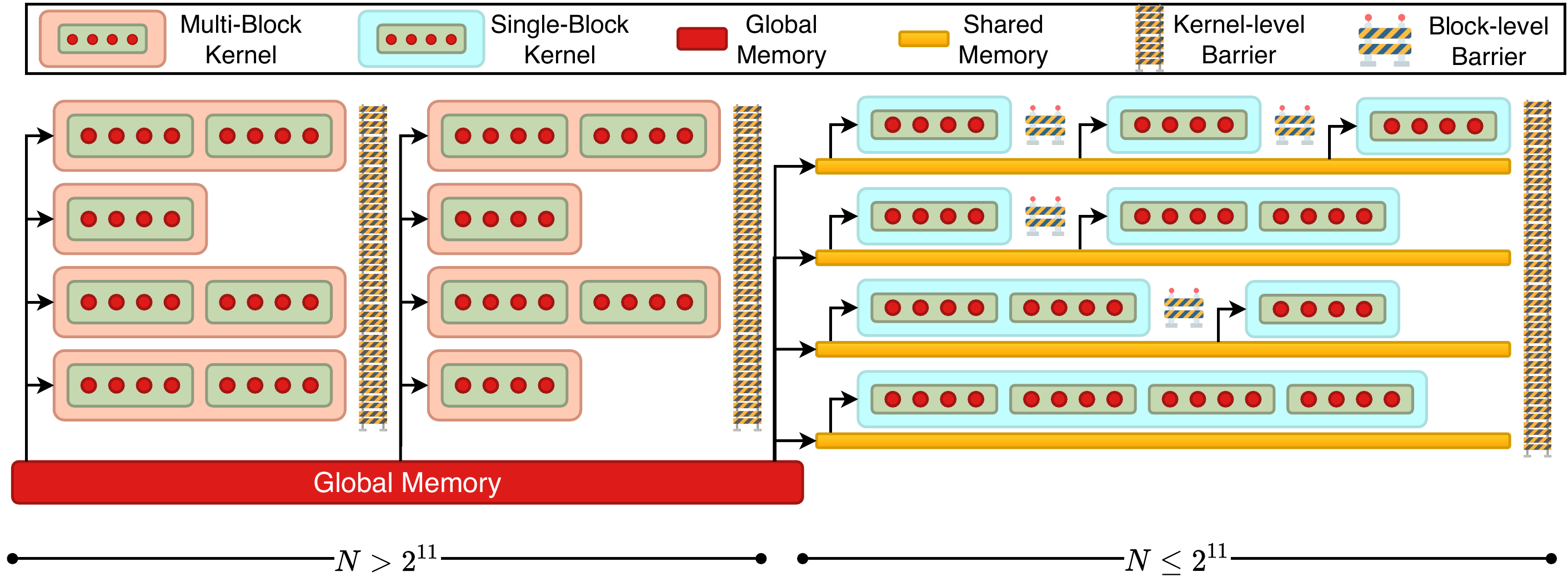}
	\vspace{-0.6em}
	\caption{The Multi-block NTT task distribution.}
	\label{fig:multi_block}
\vspace{-0.8em}
\end{figure}

The LOM-NTT kernel is designed to handle large input arrays ($N > 2^{11}$).
The LOM-NTT kernel distributes tasks using a similar strategy as used in the LOS-NTT kernel, except that it spreads them over multiple blocks.
This allows us to employ multiple SMs to execute the workload in parallel.
The LOM-NTT kernel splits a single $N$-point NTT between multiple blocks.
Because of the use of multiple blocks, this implementation requires kernel-wide barriers for synchronization between stages.
We use the LOM-NTT kernel to decompose a single $N$-point NTT into multiple $2^{11}$-point NTTs.
Then we incorporate our LOS-NTT (Single-block) kernel to evaluate all the $2^{11}$-point NTTs to harness the optimizations of shared memory and block-level barriers.
We show the distribution for our LOM-NTT for $N = 2^{16}$ in Figure~\ref{fig:multi_block}.

\subsection{Throughput-optimized Multi-block NTT}
\label{section:tom_ntt}
The throughput-optimized kernel is designed to compute multiple NTTs simultaneously.  Unlike the latency-optimized kernel that computes just a single NTT operation, TOM-NTT is optimized to compute up to $2^{15}$ NTT operations simultaneously, with each NTT computation being a $2^{16}$-point NTT (the size of each input coefficient vector is $2^{16}$).  The TOM-NTT kernel is fed $2$ input matrices.  The first matrix holds the input coefficient vectors. These vectors, of size $2^{16}$, are stacked in the matrix in the row-major format.
This matrix is then transferred to GPU and stored in global memory in a column-major format, coalescing reads across threads into a single memory transaction. 
%($1024$ threads each accessing an element from an array of $4$ byte integers).
The second input matrix contains the twiddle factors. We store twiddle factors in a similar way as the coefficient matrix.
Each input matrix is of dimension $2^{16} \times 2^{15}$. 
%With each element of size $4$ bytes, the total size of each input matrix is $2^{16} \times 2^{15} \times 4 = 8~\mathrm{GB}$.
Both matrices, when combined, completely fill the DRAM storage of $16$ GB on the V100 GPU.  The TOM-NTT kernel executes the $2^{8}$-point NTT over $32,768$ vectors in $628~\mathrm{ms}$. With an average execution time is $19.17~\mu\mathrm{s}$ per NTT operation, this kernel exhibits close to linear weak scaling.

% \begin{figure}[htbp]
% 	\centering
% 	\includegraphics[width=0.48\textwidth]{figures/06_cudantt/single_block.pdf}
% 	\caption{Single Block NTT Implementation}
% 	\label{fig:single_block}
% \end{figure}

% \input{sections/07_performance_analysis}

%  ____                 _ _       
% |  _ \ ___  ___ _   _| | |_ ___ 
% | |_) / _ \/ __| | | | | __/ __|
% |  _ <  __/\__ \ |_| | | |_\__ \
% |_| \_\___||___/\__,_|_|\__|___/

\section{Results}
% \begin{figure*}[htbp]
% 	\centering
% 	\includegraphics[width=1.0\textwidth]{figures/08_results/3_set_plot.pdf}
% % 	\vspace{-2em}
% 	\caption{(a) NTT timing comparison, varying the input coefficient vector size $N$ (b) NTT execution times for various modular reductions, with \"{O}zerk et al.~\cite{ozerk} included for comparison ($\log(\N) = 16$) (c) Polynomial multiplication time for various modular reductions (using our most optimized NTT kernel).}
% 	\label{fig:3_set_plots}
% % \vspace{-1.75em}
% \end{figure*}

% \begin{figure*}[htbp]
% 	\centering
% 	\includegraphics[width=1.0\textwidth]{figures/08_results/mod_red/mr_profile.pdf}
% % 	\vspace{-2em}
% 	\caption{Modular Reduction Arch. Profile}
% 	\label{fig:mod_red_arch_profile}
% % \vspace{-1.75em}
% \end{figure*}

% \begin{figure*}[htbp]
% 	\centering
% \includegraphics[width=1.0\textwidth]{figures/08_results/mod_red/mr_profile_stall.pdf}
% % 	\vspace{-2em}
% 	\caption{Modular Reduction Stall Profile}
% 	\label{fig:mod_red_stall_profile}
% % \vspace{-1.75em}
% \end{figure*}

% \begin{figure*}[htbp]
% 	\centering
% \includegraphics[width=1.0\textwidth]{figures/08_results/mod_red/mr_all.pdf}
% % 	\vspace{-2em}
% 	\caption{Modular Reduction Profile}
% 	\label{fig:mod_red_all}
% % \vspace{-1.75em}
% \end{figure*}

\subsection{Experimental Methodology}
We present three different NTT kernels in this work, along with four optimizations tailored for the GPU platform. We evaluate the performance of our Single-block NTT kernel for input coefficient vector sizes of $N=2^{11}$ and of our Multi-block kernel for vector sizes of $N=2^{12}~\mathrm{to}~2^{16}$.
We incrementally add each of the four optimizations to our NTT kernels and report performance improvements. Twiddle factors are pre-computed on the CPU and hence do not add to the compute overhead on the GPU. We report on multiple performance metrics for each approach, leveraging profiling tools on the GPU platform. For each optimization, the speedup achieved is reported using the respective non-optimized kernel as the baseline for comparison. Finally, we evaluate weak scaling for our throughput-optimized TOM-NTT kernel.

\subsection{Performance Metrics}
%For our experiments, we utilize NVIDIA's Volta architecture GPU, the V100 PCIe GPU with 16 GB onboard memory.
% \textcolor{blue}{ We first report the performance gain of each of our GPU optimizations for all the $3$ NTT kernel implementations. Table~\ref{table:optimizations} provides metrics for speedup obtained after incorporating our GPU optimizations in our multi-block NTT kernel (for input vector size of $N = 2^{16}$). Our multi-block NTT kernel internally utilizes our single-block NTT kernel for $2^{11}\mathrm{-}$point NTTs. Hence the metrics reported in Table~\ref{table:optimizations} are representative of both our single- and multi-block NTT kernel performances.}

We incrementally add optimizations to our NTT kernels and report performance improvements in Table~\ref{table:optimizations} (for input coefficient size $N = 2^{16}$).
For each optimization, the speedup achieved is reported, using the respective non-optimized kernel as the baseline for comparison.

\begin{table}[htb]
\vspace{-0.6em}
	\centering % used for centering table
	% p{20mm}
	\begin{tabular}{l c c c} % centered columns (4 columns)
		\hline\hline %inserts double horizontal lines
		\textbf{Optimization} & \begin{tabular}{@{}c@{}} \textbf{Relative} \\ \textbf{Speedup} \end{tabular} & \begin{tabular}{@{}c@{}} \textbf{L1\$} \\ \textbf{Throughput} \end{tabular} & \begin{tabular}{@{}c@{}} \textbf{DRAM} \\ \textbf{Throughput} \end{tabular} \\ [0.5ex] % inserts table
		%heading
		\hline %\\ [0.5ex] % inserts single horizontal line
		SM-only & $1.2\times$ & $-27.3\%$ & $+20.0\%$\\ [1ex]
		SM + Alg4 & $1.72\times$ & $+10.86\%$ & $+3.2\%$ \\ [1ex]
		SM + Alg4 + 2D & $2.91\times$ & $+5.85\%$ & $+16.14\%$ \\ [1ex]
		SM + Alg4 + 2D + FHP & $1.02\times$ & $+0.3\%$ & $-0.33\%$ \\ [1ex]
		%\hline % inserts single horizontal line
		%Fused Hadamard Product$^*$ & $1.31\times$ & $+9.9\%$ & $+18.0\%$ \\ [1ex]
		\hline \\ [1ex]  %inserts single line 
	\end{tabular}
	\caption{NTT Kernel Optimizations: SM = shared memory, Alg4 = our proposed reduction, 2D = Mixed Radix 2D NTT, FHP = Fused Hadamard Product (NTT kernel with $N=2^{16}$ and $\ceil{\log_2(q)}=62$ used as baseline) } % title of Table
	\label{table:optimizations} % is used to refer this table in the text
\vspace{-1.8em}
\end{table}

%% Shared Memory Optimization
Our shared memory optimized kernel, when compared against the global memory kernel, achieves a $20\%$ improvement in DRAM bandwidth utilization and a $1.2\times$ speedup. Data is transferred between DRAM and shared memory using coalesced memory transactions, improving DRAM bandwidth utilization.

%% Do we need to mention why L1 cache hit rate is not mentioned for Shared memory optimization in Table 2 ?
%% It is because, shared memory accesses do not hit L1 cache (Shared memory itself can be considered as a user configurable L1 cache). It is irrelevant to mention L1 cache hit rate for shared memory optimization.
%% Proposed Barrett's Reduction
Next, we compare the execution time for our NTT kernel implementation by incorporating various modular reduction techniques, as shown in Figure~\ref{fig:mod_red_all}.
% We consider the size of input polynomial as $N = 2^{16}$.
We compare our best performing NTT kernel (highlighted in Table~\ref{table:related_work}) to \"{O}zerk et al.~\cite{ozerk} and find a $1.85\times$ speedup for $N = 2^{16}$ and a $1.13\times$ speedup for $N = 2^{14}$.
% Our proposed modified Barrett's reduction reduces the number of correctional subtractions, allowing us to obtain a $1.72\times$ average speedup over classical Barrett's implementations~\cite{barrett}.
%and a $1.40\times$ average speedup over the built-in modulus operation.
%% Higher Radix Optimization
The use of radix $4$, $8$, and $16$ and $2D$ implementations provide additional speedup due to the increased temporal, as well as spatial, locality in $4$, $8$, and $16$-point butterfly operations, as compared to the baseline radix $2$ implementation. The effects of increased data locality are reflected in the $5.85\%$ improvement in the L1 cache hit-rate. Our best performing kernel, that of $2D$ NTT, achieves a $2.91\times$ speedup over a radix $2$ implementation (Figure~\ref{fig:radix_timing}).
%% Fused Hadamard Product
Our fused polynomial multiplication kernel reduced the execution time for the last stage of the merged CT NTT kernel, the Hadamard product, and the first stage of the merged GS NTT kernel,
from $8.5~\mu\mathrm{s}$ down to $6.5~\mu\mathrm{s}$, resulting in a $1.3\times$ speedup as compared to its non-fused counterpart. When incorporated within a polynomial multiplication kernel, this translates to a $6.1\%$ improvement for Single-block kernel (for size $N=2^{11}$) and a $2.4\%$ improvement for Multi-block kernel (for size $N = 2^{16}$).
% \begin{figure}[htb]
% \vspace{-0.8em}
% 	\centering
% 	\includegraphics[width=0.48\textwidth]{figures/08_results/radix_plot.pdf}
% 	\caption{NTT execution time comparison of various radix implementations.}
% 	\label{fig:radix_timing}
% \vspace{-0.7em}
% \end{figure}

% \begin{figure}[htb]
% \vspace{-0.8em}
% 	\centering
% 	\includegraphics[width=0.48\textwidth]{figures/08_results/radix/radix_profile_all.pdf}
% 	\caption{Higher Radix Comparison for NTT and inverse-NTT kernels}
% 	\label{fig:radix_timing}
% \vspace{-0.7em}
% \end{figure}

% \begin{figure}[htb]
% \vspace{-0.8em}
% 	\centering
% 	\includegraphics[width=0.48\textwidth]{figures/08_results/radix/radix_compare_ntt.pdf}
% 	\caption{NTT Radix Compare}
% 	\label{fig:radix_timing}
% \vspace{-0.7em}
% \end{figure}

% \begin{figure}[htb]
% \vspace{-0.8em}
% 	\centering
% 	\includegraphics[width=0.48\textwidth]{figures/08_results/radix/radix_compare_intt.pdf}
% 	\caption{inverse-NTT Radix Compare}
% 	\label{fig:radix_timing}
% \vspace{-0.7em}
% \end{figure}

% \begin{figure}[htb]
% \vspace{-0.8em}
% 	\centering
% 	\includegraphics[width=0.48\textwidth]{figures/08_results/test.png}
% 	\caption{Test}
% 	\label{fig:radix_timing}
% \vspace{-0.7em}
% \end{figure}

% Scaling Results
\begin{figure}[t]
% \vspace{-2.0em}
	\centering
	\includegraphics[width=0.48\textwidth]{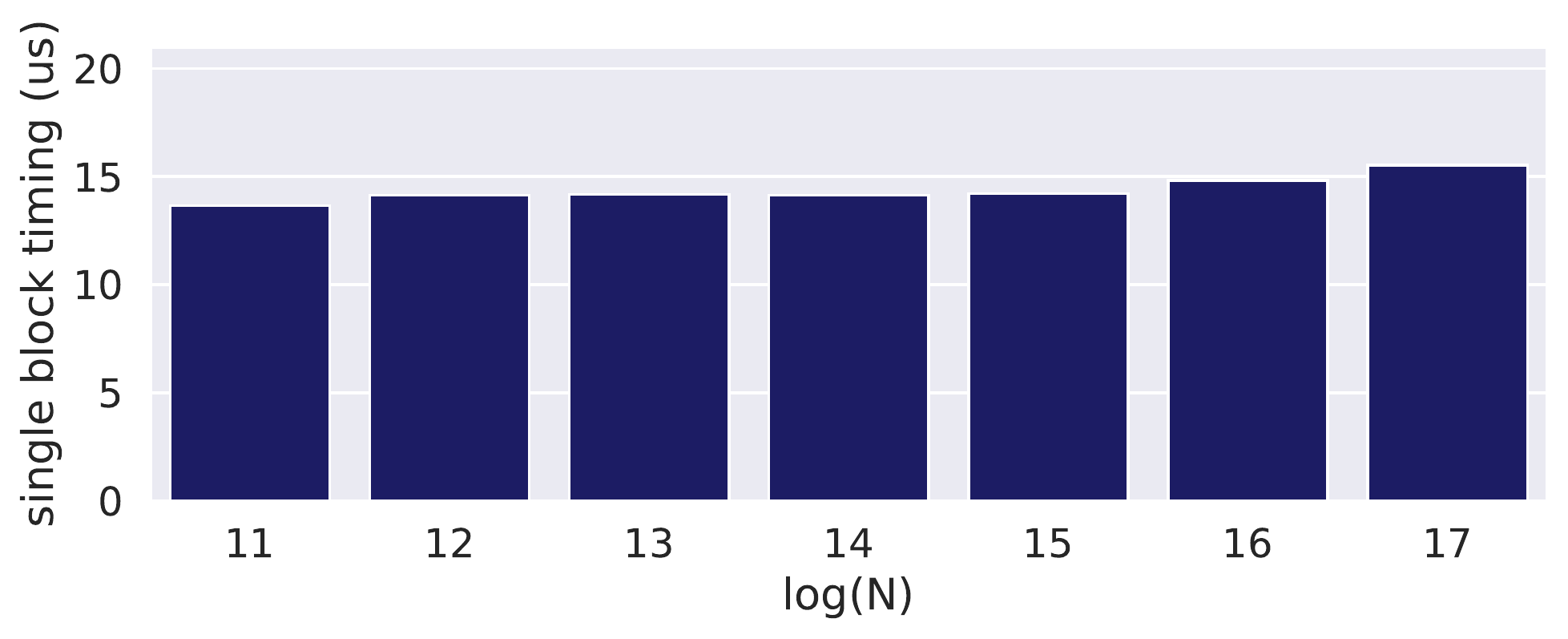}
	\vspace{-0.9em}
	\caption{Timing for the Single-block NTT.}
%\vspace{-0.9em}
	\label{fig:single_block_ntt_timing}
\vspace{-1.8em}
\end{figure}

We also measured the scalability of our fastest single-block NTT implementation. As our multi-block kernel implementation leverages our Single-block code, we also analyzed the performance of the Single-block kernel by varying the input polynomial size and the hardware resources used. On each iteration, we double the size of the input array, as well as the number of potential SMs utilized (by doubling the number of blocks in the kernel).
We observe that our Single-block kernel exhibits close to linear weak scaling, as execution times remain near constant as we increase both the input size and the hardware resources utilized (Figure~\ref{fig:single_block_ntt_timing}).

% \begin{figure}[htbp]
% 	\centering
% 	\includegraphics[width=0.48\textwidth]{figures/08_results/a.pdf}
% 	\caption{NTT timing comparison when varying the input coefficient vector size $N$.}
% 	\label{fig:ntt_comparison}
% \end{figure}
% We further incorporated these modular reduction techniques into a polynomial multiplication kernel (which includes a merged CT NTT, a merged GS NTT, and a Hadamard product kernel). The execution times for polynomial multiplication kernels (for $N = 2^{16}$) are compared in Figure~\ref{fig:mod_red_all}.
% Our best performing polynomial multiplication kernel (that is based on our proposed Barrett's modular reduction) achieves a $1.72\times$ speedup versus the the polynomial multiplication kernel that utilizes builtin modular reduction.
%(see Figure~\ref{fig:3_set_plots}c).
% To the best of our knowledge, there does not exist other $32$-bit GPU implementations that report their performance of polynomial multiplication for us to compare against.
% \begin{figure}[htbp]
% 	\centering
% 	\includegraphics[width=0.48\textwidth]{figures/08_results/b.pdf}
% 	\caption{NTT execution times for various modular reductions, with Ozerk et al.~\cite{ozerk} included for comparison ($\log(\N) = 16$).}
% 	\label{fig:ntt_compare_2}
% \end{figure}
% \begin{figure}[htbp]
% 	\centering
% 	\includegraphics[width=0.48\textwidth]{figures/08_results/c.pdf}
% 	\caption{Polynomial multiplication time comparison for various modular reductions.}
% 	\label{fig:poly_compare}
% \end{figure}

We also evaluate our TOM-NTT kernel that is optimized for operating on a large number of NTT operations simultaneously (working with up to $2^{15}$ input coefficient vectors, each of size $2^{16}$ elements).
With an average execution time of $19.17~\mu\mathrm{s}$ per NTT operation, this kernel exhibits close to linear weak scaling.
Including all optimizations, our NTT kernels achieve a speedup of $123.13\times$ and $2.37\times$ over the previous state-of-the-art CPU~\cite{sealcrypto} and GPU~\cite{ozerk} implementations of NTT kernels, respectively.

% Add roofline

% Nega-conv Kernel 2^11
%
% 1. Exec NAIVE (Builtin / without Shared memory)
% 2. Shared Mem
% 3. Fused HP + Shared Mem
% 4. Radix 4 + Fused HP + Shared Mem
% 5. Barrett + Radix 4 + Fused HP + Shared Mem

% \subsubsection{LOM-NTT}

% \subsubsection{TOM-NTT}

% 1. Analysis of modular multiplication 

% \subsection{GPU bottleneck analysis}
% \begin{enumerate}
%     \item Our implementation of NTT
    
% \end{enumerate}

% GPU work partitioning

% Comparison:
% \begin{enumerate}
%     \item Isolated benchmarks for modular multiplication of NTT
% \end{enumerate}

% Questions:
% \begin{enumerate}
%     \item Do we compare NTT against Accelerator implementations?
%     \item Comparison against CPU performance?
%     \item Comparison against cuFFT?
%     \item Do we perform isolated Modular multiplication benchmark or plug it in out NTT kernel?
%     \item 
% \end{enumerate}

% Computational graph

%  ____      _       _           _ 
% |  _ \ ___| | __ _| |_ ___  __| |
% | |_) / _ \ |/ _` | __/ _ \/ _` |
% |  _ <  __/ | (_| | ||  __/ (_| |
% |_| \_\___|_|\__,_|\__\___|\__,_|

% __        __         _    
% \ \      / /__  _ __| | __
%  \ \ /\ / / _ \| '__| |/ /
%   \ V  V / (_) | |  |   < 
%    \_/\_/ \___/|_|  |_|\_\

\section{Related Work}

Table~\ref{table:related_work} presents runtimes of various implementations of NTT and iNTT, adding to Table 8 in the work by \"Ozerk et al.~\cite{ozerk} with our own runtimes.
%Table~\ref{table:related_work} provides a comparison of NTT and iNTT workload runtimes with previous work. It borrows findings from~\cite{ozerk} and extends the comparison.
Prior studies have explored accelerated NTT on FPGAs~\cite{SadeghRiazi2020} and custom accelerators~\cite{f1}. But these custom solutions are not typically found on general-purpose systems. On the other hand, GPUs are ubiquitous and easily programmed. In recent years, there has been growing interest in using a GPU to exploit the parallelism present in NTT~\cite{zhai,kim2020,ozerk}. In particular, \"Ozerk et al.~\cite{ozerk} propose an efficient hybrid kernel approach to accelerate NTT. Our LOS-NTT and LOM-NTT kernels are inspired by their work, however, we provide some further optimizations such as our fused Hadamard product, an improved version of Barrett reduction, and explored higher radix NTTs. Kim et al.~\cite{kim2020} also propose some optimizations on NTTs, such as batching using shared memory. We explored how those optimizations could address the limitations we faced when implementing a kernel with a radix higher than 4.

Alkim et al.~\cite{alkim} define and analyze several algorithms very similar to Algorithm~\ref{alg:fusedhadamard}. They not only consider truncating their NTTs by one stage but by two and three stages. Although some of Alkim et al.'s algorithms utilize Karatsuba's Algorithm, they do not consider using Karatsuba's Algorithm to merge a single innermost pair of NTT stages. In our tests, our fused polynomial multiplication implementation provides an additional speedup of $6.1\%$ and $2.4\%$ as compared to the naive implementation for polynomial multiplication for input sizes of $N = 2^{11}$ and $N = 2^{16}$, respectively using Alkim et al.'s $(k-1)$-level NTT multiplication algorithm.

\begin{table}[t]
\setlength{\tabcolsep}{5pt} % Default value: 6pt
\vspace{-0.6em}
	\centering % used for centering table
	% p{20mm}
	\begin{tabular}{l l c c c c} % centered columns (4 columns)
		\hline\hline %inserts double horizontal lines
		\textbf{Work} & \textbf{Platform} & $N$ & $\ceil{\log_2(q)}$ & \begin{tabular}{@{}l@{}} \textbf{NTT} \\ $(\mu s)$ \end{tabular} & \begin{tabular}{@{}l@{}} \textbf{iNTT} \\ $(\mu s)$ \end{tabular} \\ [1ex]
		%heading
		% 2^14 = 16384
		% 2^15 = 32768
		% 2^16 = 65536
		\hline %\\ [0.5ex] % inserts single horizontal line
		cuHE~\cite{cuhe}$^*$ & GTX $690$ & $2^{14}$  &  $64^c$ & $56$ & $65.3$ \\ [1ex]
		 & & $2^{15}$  &  $64^c$ & $71.2$ & $83.6$ \\ [1ex]
		cuHE~\cite{cuhe}$^{*,a}$ & Tesla $K80$ & $2^{14}$  &  $64^c$ & $12.9$ & $12.5$ \\ [1ex]
		& & $2^{15}$  &  $64^c$ & $19$ & $21.6$ \\ [1ex]
		cuHE~\cite{cuhe}$^{*,b}$ & GTX $1070$ & $2^{14}$  &  $64^c$ & $66.8$ & $-$ \\ [1ex]
		Faster NTT~\cite{al2018faster}$^*$ & Tesla $K80$ & $2^{14}$  &  $64^c$ & $9.6$ & $9.7$ \\ [1ex]
		& & $2^{15}$  &  $64^c$ & $15.3$ & $16.2$ \\ [1ex]
		Accl NTT~\cite{goey}$^*$ & GTX $1070$ & $2^{14}$  &  $64^c$ & $57.8$ & $-$ \\ [1ex]
		% Cloud NTT~\cite{zheng2020encrypted} & RTX $2080$Ti & $2^{15}$  &  $24$ & $83.3$ & $96$ \\ [1ex]
		Bootstrap HE~\cite{kim2020} & Titan $V$ & $2^{14}$  &  $60$ & $44.1$ & $-$ \\ [1ex]
		 & & $2^{15}$  &  $60$ & $84.2$ & $-$ \\ [1ex]
		Re-encrypt~\cite{sahu} & GTX $1050$ & $2^{14}$  &  NA & $255$ & $-$ \\ [1ex]
		 & & $2^{15}$  &  NA & $470$ & $-$ \\ [1ex]
		 & RTX $1080$ & $2^{14}$  &  NA & $375$ & $-$ \\ [1ex]
		 & & $2^{15}$  &  NA & $425$ & $-$ \\ [1ex]
		Efficient NTT~\cite{ozerk} & GTX $980$ & $2^{14}$  &  $55$ & $51$ & $41$ \\ [1ex]
		 & & $2^{15}$  &  $55$ & $73$ & $52$ \\ [1ex]
		 & GTX $1080$ & $2^{14}$  &  $55$ & $33$ & $20$ \\ [1ex]
		 & & $2^{15}$  &  $55$ & $36$ & $24$ \\ [1ex]
		& Tesla $V100$ & $2^{14}$  &  $55$ & $29$ & $21$ \\ [1ex]
		 & & $2^{15}$  &  $55$ & $39$ & $23$ \\ [1ex]
% 		Our Work & Tesla $K80$ & $2^{14}$  &  $32$ & $-$ & $-$ \\ [1ex]
% 		 & & $2^{15}$  &  $32$ & $-$ & $-$ \\ [1ex]
% 		& GTX $1080$Ti & $2^{14}$  &  $32$ & $-$ & $-$ \\ [1ex]
% 		 & & $2^{15}$  &  $32$ & $-$ & $-$ \\ [1ex]
		Our Work & Tesla $A100$ & $2^{14}$  &  $62$ & $13.3$ & $10.9$ \\ [1ex]
		                      & & $2^{16}$  &  $62$ & $16.5$ & $18.7$ \\ [1ex]
		         & Tesla $V100$ & $2^{14}$  &  $30$ & $8.7$  & $10.0$ \\ [1ex]
		                      & & $2^{16}$  &  $30$ & $13.1$ & $13.4$ \\ [1ex]
		                      & & $2^{14}$  &  $62$ & $11.5$ & $11.9$ \\ [1ex]
		                      & & $2^{16}$  &  $62$ & $\boldsymbol{16.4}$ & $\boldsymbol{17.3}$ \\ [1ex]
		
		\hline  %inserts single line 
	\end{tabular}
	\begin{flushleft}
	\begin{tabular}{l}
 	$^*$uses constant prime $q=\texttt{0xFFFFFFFF00000001}$ \\
	%$^*$These implementations fix a single 64-bit prime \\ $q=\texttt{0xFFFFFFFF00000001}$, known as a Solinas prime \\
	$^a$results are from~\cite{al2018faster} \hspace{8mm}$^b$results are from~\cite{goey} \\
	$^c$actual $q_i$ is restricted by $q_i^2n<2^{64}-2^{32}+1$
	\end{tabular}
	\end{flushleft}
	\caption{Comparison to related work} % title of Table
	\label{table:related_work} % is used to refer this table in the text
\vspace{-3.3em}
\end{table}

There is a Barrett reduction variant proposed by Yu et al.~\cite{yu} that requires no correctional subtractions. We found that this algorithm has severe trade-offs in terms of operational complexity as a function of workload size, which makes it less attractive for use with HE.

\section{Conclusion}
In this work, we presented an analysis and proposed implementations of polynomial multiplication, the key computational bottleneck in lattice-based HE systems, while targeting the V100 GPU platform. Specifically, we analyzed Barrett's modular reduction algorithm and several variants. We studied the interplay between algorithmic improvements (such as multi-radix NTTs) and low-level kernel optimizations tailored towards the GPU (including memory coalescing). 
% Our proposals provide a $1.85\times$ speedup over the state-of-the-art.
Our NTT optimizations achieve an overall speedup of $123.13\times$ and $2.37\times$ over the previous state-of-the-art CPU~\cite{sealcrypto} and GPU~\cite{ozerk} implementations of NTT kernels, respectively.

% In this work, we presented a thorough analysis of polynomial multiplication, an optimized implementation of this kernel targeting the NVIDIA V100 GPU.
% %% Barrett's reduction
% We further explore optimization of modular reduction techniques, analyzing previous attempts to accelerate the Barrett reduction algorithm and developed our own optimized algorithm that provided $XX\times$ speedup over classical Barrett's reduction and $XX\times$ speedup over built-in modular reduction.

%% Is this statement necessary????
%We believe our contributions will promote the development of future FHE workloads and raise awareness in both the cryptology community and the architecture industry, emphasizing the need for accelerating polynomial multiplication for FHE.

%     _    ____ _  __
%    / \  / ___| |/ /
%   / _ \| |   | ' / 
%  / ___ \ |___| . \ 
% /_/   \_\____|_|\_\

\section*{Acknowledgements}

This work was supported in part by 
the Institute for Experiential AI, 
the Harold Alfond Foundation, 
the NSF IUCRC Center for Hardware and Embedded Systems Security and Trust (CHEST),
the RedHat Collaboratory, 
and project grant PID2020-112827GB-I00 funded by MCIN/AEI/10.13039/501100011033.
\balance
\bibliographystyle{IEEEtran}
%\balance
\bibliography{bibliography/biblio}

\end{document}